\documentclass[12pt]{article}
\pdfoutput=1
\usepackage{amsmath,amssymb,graphicx} %use drftcite in draftmode
\usepackage{epsf}
\usepackage{pstricks}
\usepackage{cite}

\newcommand{\beq}{\begin{eqnarray}}% can be used as {equation} or  {eqnarray}
\newcommand{\eeq}{\end{eqnarray}}

\newcommand{\centeron}[2]{{\setbox0=\hbox{#1}\setbox1=\hbox{#2}\ifdim
\wd1>\wd0\kern.5\wd1\kern-.5\wd0\fi
\copy0

\kern-.5\wd0\kern-.5\wd1\copy1\ifdim\wd0>\wd1
                                       \kern.5\wd0\kern-.5\wd1\fi}}
\newcommand{\ltap}{\>\centeron{\raise.35ex\hbox{$<$}}
                               {\lower.65ex\hbox{$\sim$}}\>}
\newcommand{\gtap}{\>\centeron{\raise.35ex\hbox{$>$}}
                               {\lower.65ex\hbox{$\sim$}}\>}

\newcommand{\gtsimeq}{\raisebox{-0.6ex}{$\,\stackrel
        {\raisebox{-.2ex}{$\textstyle >$}}{\sim}\,$}}

\newcommand\ZZ{\hbox{\zfont Z\kern-.4emZ}}
\font\zfont = cmss10 %scaled \magstep1

\begin{document}
\begin{titlepage}
\begin{flushright}
\end{flushright}

\vskip.5cm
\begin{center}
{\huge \bf
The AdS/QCD Correspondence: \\
\vskip0.2cm
Still Undelivered
}

\vskip.1cm
\end{center}
\vskip0.2cm

\begin{center}
{\bf
 {Csaba Cs\'aki}$^a$, {Matthew Reece}$^b$,
{\rm and}
{John Terning}$^c$}
\end{center}
\vskip 8pt

\begin{center}

$^{a}$ {\it Institute for High Energy Phenomenology\\
Newman Laboratory of Elementary Particle Physics\\
Cornell University, Ithaca, NY 14853, USA } \\

\vspace*{0.2cm}

$^{b}$ {\it Princeton Center for Theoretical Science\\
Princeton University, Princeton, NJ 08544, USA}

\vspace*{0.2cm}

$^{c}$ {\it
Department of Physics, University of California, Davis, CA
95616.} \\

\vspace*{0.1cm} \vspace*{0.3cm} {\tt  csaki@lepp.cornell.edu,
mreece@princeton.edu, jterning@gmail.com}
\end{center}

\vglue 0.3truecm

\begin{abstract}
\vskip 3pt \noindent We consider the particle spectrum and event
shapes in large $N$ gauge theories in different regimes of the
short-distance 't Hooft coupling, $\lambda$. The mesons  in the
small $\lambda$ limit should have a Regge spectrum in order to agree
with perturbation theory, while generically the large $\lambda$
theories with gravity duals produce spectra reminiscent of KK modes.
We argue that these KK-like states are qualitatively different from
QCD modes: they are deeply bound states which are sensitive to short
distance interactions rather than the flux tube-like states expected
in asymptotically free, confining gauge theories. In addition, we
also find that the characteristic event shapes for the large
$\lambda$ theories with gravity duals are close to spherical, very
different from  QCD-like (small $\lambda$, small $N$) and
Nambu-Goto-like (small $\lambda$, large $N$) theories which have
jets. This observation is in agreement with the conjecture of
Strassler on event shapes in large 't Hooft coupling theories, which
was recently proved by Hofman and Maldacena for the conformal case.
This conclusion does not change even when considering soft-wall
backgrounds for the gravity dual. The picture that emerges is the
following: theories with small and large $\lambda$ are qualitatively
different, while theories with small and large $N$ are qualitatively
similar. Thus it seems that it is the relative smallness of the 't
Hooft coupling in QCD that prevents a reliable AdS/QCD
correspondence from emerging, and that reproducing characteristic
QCD-like behavior will require genuine stringy dynamics to be
incorporated into any putative dual theory.
\end{abstract}

\end{titlepage}

\newpage

%\renewcommand{\thefootnote}{(\arabic{footnote})}

%%%%%%%%%%%%%%%%%%%%%%%%%%%%%%%%%%%%%%%%%%%%%%%%%%%%%%
%%%%%%%%%%%%%%%%%%%%%%%%%%%%%%%%%%%%%%%%%%%%%%%%%%%%%%
\section{Introduction}
\label{sec:intro}
\setcounter{equation}{0}
\setcounter{footnote}{0}
%%%%%%%%%%%%%%%%%%%%%%%%%%%%%%%%%%%%%%%%%%%%%%%%%%%%%%
%%%%%%%%%%%%%%%%%%%%%%%%%%%%%%%%%%%%%%%%%%%%%%%%%%%%%%

Confining gauge theories play a central role in particle physics:
they provide the real theory of strong interactions (QCD), and
possible models of dynamical electroweak symmetry breaking (EWSB)
and dynamical supersymmetry breaking. With the help of the anti-de
Sitter/conformal field theory (AdS/CFT) correspondence many of the
qualitative features of confining gauge theories can be explained,
and warped extra dimensional theories have been used to parametrize
strongly coupled, almost conformal, models of dynamical EWSB. The
AdS/CFT analogy \cite{maldacena} has been pushed so far as to
suggest that most properties of QCD can be understood even
quantitatively using a simple Randall-Sundrum (RS)-type extra
dimensional model called AdS/QCD: a warped extra dimension with UV
and IR cut-off branes, and bulk fields appropriate for the chiral
symmetries of QCD \cite{AdSQCD}. While many results show a
surprisingly good agreement (${\cal O}(5-10)\% $) with the QCD data,
it always has been a central question whether this agreement is
merely the result of properly incorporating chiral symmetries, or if
there is some deeper, underlying reason. One recent development that
suggests that AdS/QCD is actually not that similar to real QCD is
the conjecture of Strassler that event shapes in theories at large
't Hooft coupling are spherical \cite{Strassler:2008bv}. This
possibility was already strongly suggested by the result of
Polchinski and Strassler that in hadrons at large 't Hooft coupling,
all partons are at small-$x$ (``wee")~\cite{PolchinskiStrasslerDIS};
thus a parton shower with highly energetic partons is very unlikely.
The spherical-event conjecture was proven by Hofman and Maldacena,
who in a pioneering work set out to understand the collider physics
of conformal theories with a gravity dual \cite{HofmanMaldacena}.
What they found confirmed that these theories behave very
differently from QCD: instead of jet-like events one finds energies
that are distributed in a spherically symmetric way in the
calorimeter. For related work see~\cite{HIM} and
also~\cite{MoreHatta}. One aim of this paper is to verify that the
results of Hofman and Maldacena persist in the case of a theory with
conformal symmetry broken in the IR as in the case of AdS/QCD.
Indeed we find that introducing the IR cutoff does not substantially
modify the basic results, and generic scattering processes in
AdS/QCD-like theories will lead to events with large sphericities
(as opposed to jet-like events in QCD). Of course this result is not
the first qualitative disagreement between AdS/QCD and real QCD: the
resonance spectrum of a simple extra dimensional model is also quite
different from QCD, unless a soft-wall metric is assumed for the 5D
background. We have checked that the spherical nature of the event
shapes persists even when the simple RS background is replaced by
the soft-wall background. These results suggest that there is a real
disagreement between theories with a gravity dual and QCD, as one
would have expected from the beginning since gravity duals are only
calculable in theories with large 't Hooft coupling.

\begin{figure}[h]
\begin{center}
\includegraphics[width=7.0cm]{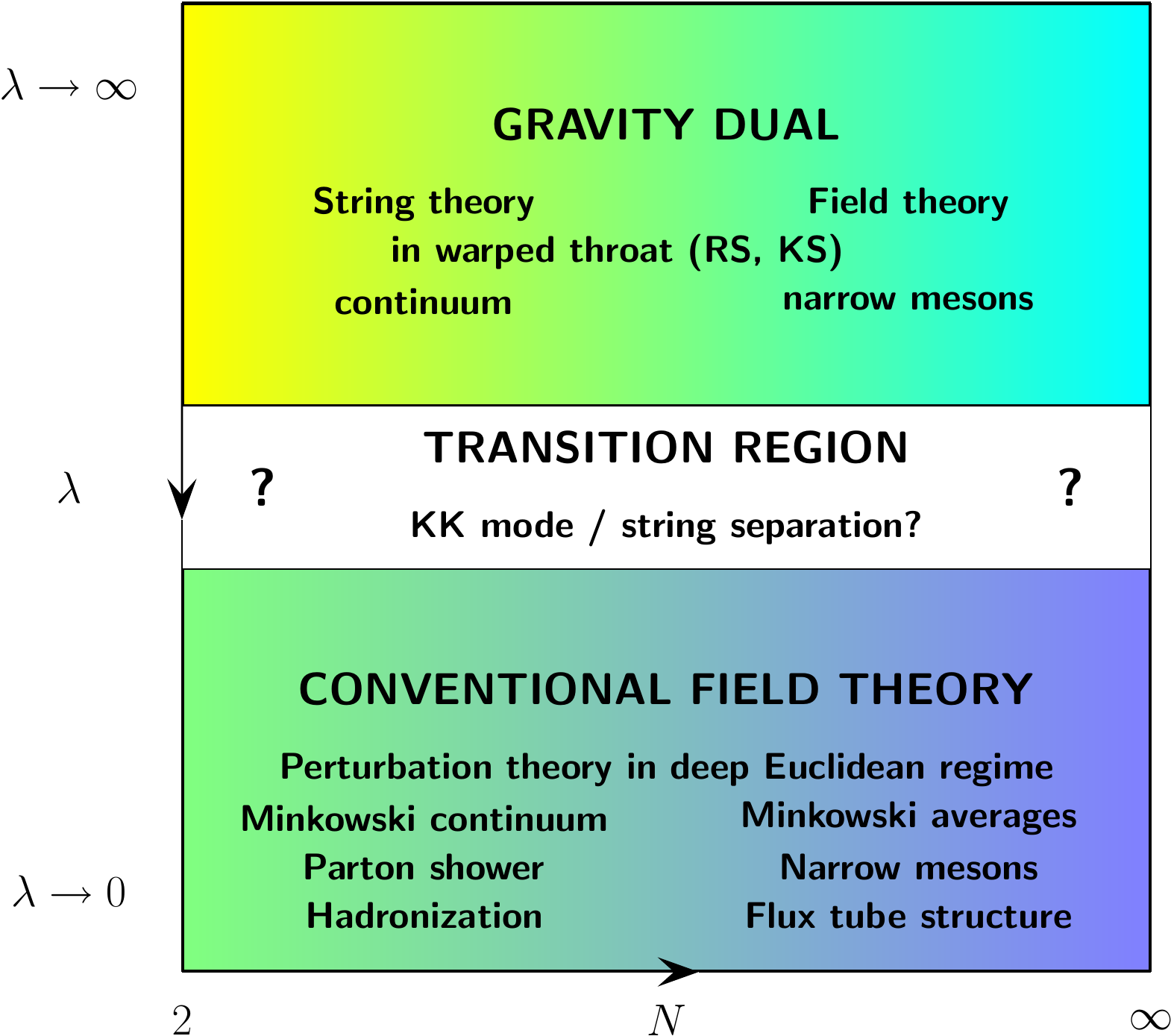}
\end{center} \caption{A survey of the $N -\lambda$ plane.} \label{fig:plane}
\end{figure}

In order to better understand what these differences can be let us
be more specific and consider QCD as a function of the number of
colors $N$ and the 't Hooft coupling $\lambda = g^2 N$. In
particular, we are concerned mostly with the value of $\lambda$ at
distances much shorter than the confinement scale, so that by
``small $\lambda$" we mean asymptotically free or at least
asymptotically weak coupling (e.g. a theory with a Banks-Zaks fixed
point in the UV). A brief survey of the $(N,\lambda )$ plane is
illustrated in Fig.~\ref{fig:plane}. At large $N$, and for any
$\lambda$, we expect a theory with some stringy reorganization in a
$1/N$ expansion \cite{'tHooft:1973jz} (but not necessarily a gravity
dual). On the other hand, for large $\lambda$ we expect the
existence of a weakly coupled gravity dual, which is generically a
string theory; however for large $N$ the string modes decouple and a
pure gravity dual should be enough to describe the theory. Thus we
get the basic characterization of the four corners of the
$(N,\lambda)$ plane: for large $N,\lambda$ the theory is reminiscent
of a Randall-Sundrum \cite{RS} (or Klebanov-Strassler \cite{KS})
model: field theory (and classical gravity) on a warped background.
For large $\lambda$, small $N$ one should find string theory on a
warped background. For large $N$, small $\lambda$ we still expect a
stringy description, but not necessarily higher dimensional, while
for small $N, \lambda$ we get ordinary perturbative QCD, with its
characteristic parton showers followed by hadronization.

The main aim of this paper is to examine the various transitions in
the $(N,\lambda)$ plane. At first sight changing $N$ from small to large
values looks like a very dramatic transition,
since in the large $N$ limit the decays of mesons will be
$1/N$-suppressed. As a result, at strictly infinite $N$, the
theory behaves vastly differently from a small $N$ field theory.
In infinite $N$ QCD, an $e^+ e^-$ collider would produce no
hadrons at all, except when tuned to have {\em precisely} the energy
of a resonance.  However
at small enough $N$, above the first few resonances, the spectral
function will be dominated by a smooth, flat continuum. This is true
at any $\lambda$: the cross section at small $N$ is continuum, but
at large $N$ is a discretuum. At very large but finite $N$, we
can study production of narrow on-shell mesons which subsequently decay.
In the large-$\lambda$ case
these mesons are KK modes. In the small-$\lambda$ case,
they are flux tubes.

Arguments for the flux tube description of hadrons date back to the
1970s, and include in particular 't Hooft's characterization of
phases of gauge theories based on the behavior of Wilson and 't
Hooft operators \cite{'tHooft:1977hy}. Roughly speaking, there is a
discrete set of possibilities for these behaviors, one of which is
the confining area-law behavior of the Wilson loop, which suggests
that the partition function has contributions of the form
$e^{-\Lambda^2 L T}$. This suggests that there are states in the
theory with constant energy density per unit length, so that their
energy is $E = \Lambda^2 L$. Such states would look stringy.
Various aspects of this argument have been made with various
amounts of precision over many years; a partial list of references
can be found in Ref. \cite{confinement}. The upshot is that confinement,
in the sense of area-law behavior over some range of distances,
implies that hadrons are approximately
described by the Nambu-Goto action. There are
similar flux tube states at large 't Hooft coupling
\cite{stringyRegge,PRT}, but they are parametrically heavier, and we
will not discuss them here.

Our goal in this paper is to argue that the small $N$ to large
$N$ transition, at given $\lambda$, is fairly smooth and does not
drastically alter qualitative properties of the physics, despite the
singular nature of the infinite $N$ limit; in other words, we
expect to have a significant overlap between partonic, QCD-like
descriptions, and hadronic, narrow-resonance descriptions. This
supports the picture of ``parton-hadron duality." On the
other hand, the transition from small to large $\lambda$ is in some
ways more mysterious. Large 't Hooft  coupling produces deeply bound
meson (and glueball) states that do not resemble the experimentally observed mesons
(in particular, they do not fall on linear Regge trajectories), but
which are calculable in supergravity duals
\cite{WittenBH, KKmodes, MesonSpect, quarkonium5D, AdSMesons,KSKK}.
The exotic properties of light, low-spin states at large 't Hooft
coupling have been emphasized in Refs. \cite{PolchinskiStrasslerDIS, 
MesonSpect, quarkonium5D}. Certain decay modes and form factors 
are strongly suppressed at large 't Hooft coupling \cite{deconstructed, quarkonium5D}.
Other aspects of the large 't Hooft coupling physics have been
studied that contribute to the picture that strong coupling at
short distances leads to novel physics \cite{annulons, adjointtrap, PlasmaBalls}.
The most recent such observation is that, whereas small
$\lambda$ leads to the familiar jet structure in QCD-like events,
large $\lambda$ leads to very spherical events as proved by
Hofman and Maldacena. As far as we are aware, there is no tool
currently available that can tell us whether  the spectrum and event
shapes transition smoothly as a function of $\lambda$ or change
abruptly at some critical $\lambda$. The latter possibility may seem
strange, but a similar transition happens in the spectrum of modes
of a static flux tube in ${\cal N}=4$ SUSY Yang-Mills \cite{N4FluxTubes}. At large
$\lambda$ many issues involving very high energy scattering of hadrons,
such as Regge trajectories, the Pomeron, and saturation
\cite{PolchinskiStrassler, PolchinskiStrasslerDIS, Eikonal}, have been studied.
Despite substantial quantitative differences, and some qualitative ones (e.g. the Pomeron
couples to individual partons at small $\lambda$, but to hadrons at large $\lambda$),
many aspects of the physics of high energy scattering 
(e.g. the saturation of the Froissart bound) are similar
at small and large $\lambda$. Nonetheless, the details of dynamical
questions at energies near the scale of the lightest resonances
are strikingly different at small and large $\lambda$.\footnote{Another 
argument against the existence of a gravity dual of 
QCD was recently made in \cite{Unsal}.} We should
emphasize that this idea occurs, in various forms, throughout
the literature, and is not new to us. However, we hope to present
particularly simple and clear illustrations of the different
regimes and make it apparent that these differences are generic,
not artifacts of particular examples that have been studied in the past.

The paper is organized as follows: in section~\ref{sec:mero} we
study the connections between the predictions of perturbation theory (PT)
for a two-point function  of currents, $\langle J(x) J(0)\rangle$,
and the large $N$ spectrum. At any $\lambda$, the deep Euclidean
behavior of such a two-point function is determined by conformal
symmetry, and unitarity relates the Minkowski and Euclidean limits.
Various possible spectra are consistent with these two facts.
However, in the perturbative case, we expect the leading-order deep
Euclidean calculation to capture the physics almost everywhere in
the complex plane, with small corrections. We confirm this
expectation by showing that a large-$N$ Regge-type spectrum, $m_n^2
\sim n$, is consistent with PT  in most of the
complex plane, except very near the Minkowski axis. At finite $N$
the poles become cuts, but the behavior of the two-point function in
most of the complex plane will be similar. On the other hand, the
RS-type spectrum, $m_n \sim n$, deviates sharply from PT
in much of the complex plane. There is nothing wrong with
this, but it implies we should only expect such a spectrum in
theories with large 't Hooft coupling, where PT
breaks down very badly. The analysis of the spectrum continues in
Section \ref{sec:toyspectrum}, which presents a simple
quantum-mechanical toy model that illustrates how the spectrum of
light states can change continuously from flux tube-type to RS-type
as a coupling constant is varied. This suggests that the KK-like
modes that are characteristic to AdS/QCD are actually not QCD
states, but rather they are deeply bound mesons, that are more
related to the short-distance Coulomb-like potential, and not to the
long-distance confining linear potential.

Having established the large-$N$ spectrum looks quite different at
large and small $\lambda$, we turn our attention to more detailed
properties of scattering events. We look at large but finite $N$, so that we
can produce a narrow resonance on-shell and watch it decay in a
sequence of steps. First we examine decays of Kaluza-Klein modes in
an extra dimension in section \ref{sec:KKmodes}. We show that the
end result of the decay chain is a large number of the lightest
modes moving in approximately independent directions, so that
events appear spherical. We repeat the analysis for a soft-wall type
background, and show that the spherical nature of the event shapes
remains. Although we cannot reliably calculate the same process at
small $N$, where stringy corrections would be important, we note
that the physics is similar to that computed by Hofman and
Maldacena, in which conformal theories at large $N$ and $\lambda$
give rise to spherical events. We speculate that large sphericities
are characteristic of any large $\lambda$ theory.

In section \ref{sec:fluxtubes}, we repeat this analysis for a different model
of hadrons, namely quasi-stable flux tubes as we would expect to find in
small $\lambda$, large $N$ theories. We study models of flux tube
evolution and breaking (motivated by Schwinger pair-production in
the chromoelectric field), and find that the resulting events have a characteristic
``jetty" structure, with energetic particles moving in opposite directions and only
relatively soft particles in between. This is very similar to the well-known jet
structure that arises from perturbative QCD and the parton shower, independently
of any assumptions about hadronization. We take this as evidence that the
small-$N$/large-$N$ transition at small $\lambda$ is much smoother
than the small to large $\lambda$ transition at large $N$.

These pieces of evidence, taken together, provide support for our picture
of the $(N, \lambda)$ plane. The old idea of parton-hadron duality is
plausibly true everywhere in that plane, with the main effect of varying $N$
being to broaden mesons until they merge with a continuum, with a
large region of overlap between partonic and hadronic descriptions. The
effects of raising $\lambda$ are much more dramatic, with the spectrum
of light excitations and the shape of events in a detector being starkly
different in the small and large $\lambda$ limits. We have no information
about what the transition region is like; it may be smooth, with no phase
transition, but it is at least a crossover between very different regimes. This
suggests that if AdS/QCD is ever to be useful for truly QCD-like theories,
we will need qualitatively new ideas and a better understanding of this
transition region.

%%%%%%%%%%%%%%%%%%%%%%%%%%%%%%%%%%%%%%%%%%%%%%%%%%%%%%
%%%%%%%%%%%%%%%%%%%%%%%%%%%%%%%%%%%%%%%%%%%%%%%%%%%%%%
\section{Perturbation Theory and the Large-$N$ Spectrum}
\label{sec:mero}
\setcounter{equation}{0}
\setcounter{footnote}{0}
%%%%%%%%%%%%%%%%%%%%%%%%%%%%%%%%%%%%%%%%%%%%%%%%%%%%%%
%%%%%%%%%%%%%%%%%%%%%%%%%%%%%%%%%%%%%%%%%%%%%%%%%%%%%%

We are interested in understanding  qualitative properties of
confining gauge theories in the plane of the number of colors $N$
and the 't Hooft coupling $\lambda$. Because we are interested in
qualitative properties, we won't concern ourselves too much with
which theory we are working in. In particular, for QCD it does not
make sense to take $\lambda$ very large, unless one considers a
brane construction of a QCD-like theory with new degrees of freedom
as in Ref. \cite{SakaiSugimoto}. For concreteness, one could imagine
studying the ${\cal N}=1^*$ theory obtained by mass deformation of
${\cal N}=4$ super-Yang-Mills, which has confining vacua in which
one can vary $N$ and $\lambda$ \cite{N1Star, PS1}. In any case, we
never have a fully controlled calculation over the whole plane, so
we will have to pick certain corners and study them. In this paper
we will be content with simplified calculations that exhibit general
qualitative properties of the corners of the $(N, \lambda)$ plane
that we are interested in.

In a theory with small 't Hooft coupling at short distances, we can
do a reliable perturbative calculation of the  two-point function of
a current in the deep Euclidean region. We trust from the operator
product expansion that corrections to the calculated behavior are
controlled: they are either proportional to a small coupling, or
they are power-suppressed. The total cross section of a process like
$e^+ e^- \rightarrow$ hadrons is computed by analytically continuing
the Euclidean answer to the Minkowski region and taking the
discontinuity across a cut. At large $N$, we know that the cross
section vanishes except exactly on a resonance. This tells us that
the large-$N$ behavior of the two-point function is controlled by
a meromorphic function, i.e. one which is analytic throughout the
complex plane except at a discrete (but infinite) set of poles. It
turns out that there are different meromorphic functions that have
the same deep Euclidean asymptotics. However, we will argue that
some of these meromorphic functions are much closer to the
perturbative answer over the whole complex plane than the
AdS/QCD-like meromorphic function, so already PT
for the two-point function seems to be telling us something
important about the spectrum of hadrons.

\begin{figure}[h]
\begin{center}
\includegraphics[width=6.5cm]{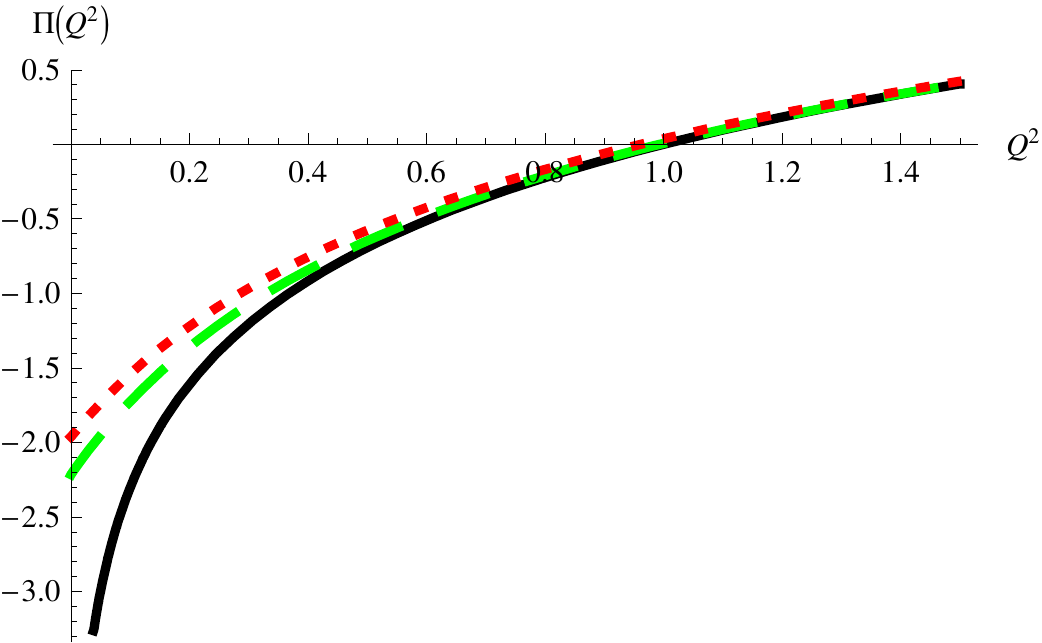}
\end{center}
\caption{Perturbation theory vs. meromorphization at Euclidean momentum, $\Pi(Q^2)$:
black solid line, $\log Q^2$; green dashed line, Pad\'e approximation;
red dotted line, digamma function.}
\label{fig:eucl}
\end{figure}

Two decades before AdS/CFT or any examples of confinement with large 't Hooft coupling
at short distances,
Migdal attempted to find a large $N$ ansatz for the QCD
spectrum by Pad\'e approximation of perturbative Euclidean
calculations \cite{Migdal77}. The idea behind the Pad\'e approximation is to produce an analytic
continuation of the perturbative result for $\Pi (Q^2)$ into the
non-perturbative regime. The leading order perturbative result for the two-point function of two conserved currents is
just given by (dropping a factor of $\frac{N}{12 \pi^2}$ for
convenience)
\begin{equation}
\Pi_{PT} (s) =\log \frac{s}{\mu^2} +{\cal O} \left(\frac{1}{s^2}\right),
\end{equation}
with $s=Q^2$. In the Pad\'e approximation one is looking for a ratio
of two polynomials $\Pi_P (s)= P_N(s)/Q_N(s)$ which at a given point
(for example $s=1$) reproduces the first $2N+1$ terms in the Taylor
series of $\Pi_{PT}$. One can then take a limit as the order of these polynomials increases,
finding a function with infinitely many poles.
The resulting function obtained this way for
$N\to \infty$ is
\begin{equation}\label{eq:Pade}
\Pi_P(s) = \frac{J_0(\sqrt{-s/\beta}) \log(-s/\beta) - \pi
Y_0(\sqrt{-s/\beta})}{J_0(\sqrt{-s/\beta})}.
\end{equation}
Here $\beta$ is chosen as $(2 x_1^2)^{-1}$ where $x_1$ is the smallest root
of $J_0(x)$; this is chosen  such that $\Pi_P(s)$ has its first pole
at $s = 1/2$. We can see in Fig.~\ref{fig:eucl} that it is a very
good approximation to the PT  result for large
Euclidean momenta (by construction).

\begin{figure}[!h]
\begin{center}
\includegraphics[width=6.5cm]{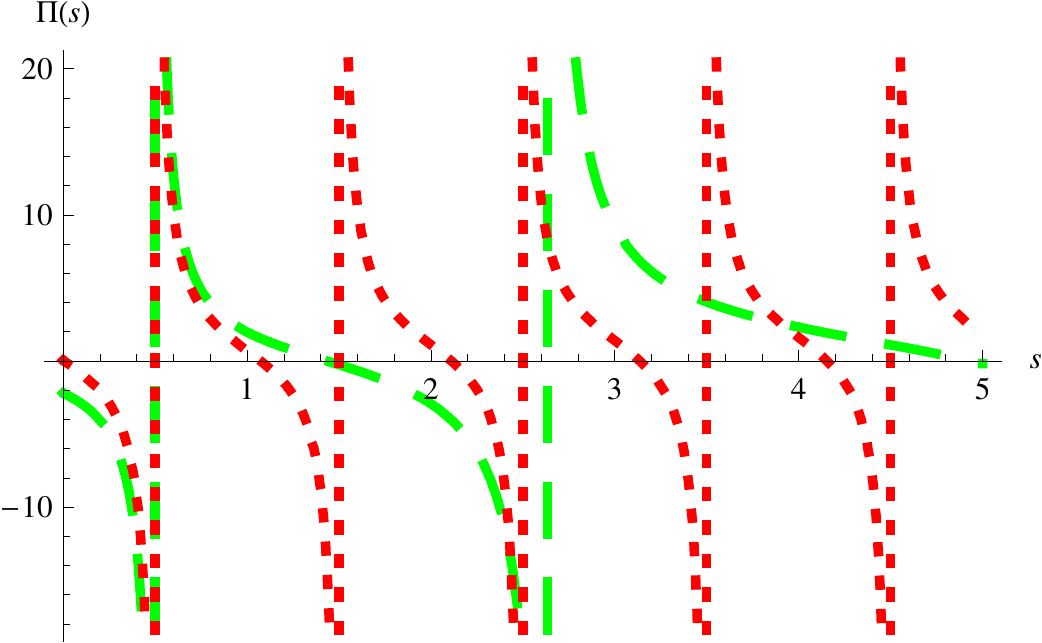}
\end{center}
\caption{Different mass spectra with similar deep Euclidean asymptotics for $\Pi(s)$:
green dashed line, analytic continuation of the Pad\'e approximation;
red dotted line, digamma function.}
\label{fig:mink}
\end{figure}

It is well-known that
this Pad\'e approximation fails to capture some of the essential
features of the non-perturbative physics of QCD. For example, the poles of Eq. (\ref{eq:Pade})
appear at $m_n^2 \sim n^2$, in stark contrast with the expected
Regge answer $m_n^2 \sim n$. The difference between two-point functions
with these types of behavior in the Minkowski region is illustrated in Figure \ref{fig:mink}.
While Migdal's answer does not match the expected QCD-like behavior,
it is remarkably precisely what one would calculate in the AdS/QCD
framework \cite{AdSQCD}, which can be considered as a simple model
of what this quantity would look like in a theory at large 't Hooft coupling.
This correspondence was first pointed out by Shifman in Ref. \cite{Shifman05},
and has been further discussed in Refs. \cite{emerging}. Thus, while
Migdal failed to produce a good model of the QCD two-point function,
he stumbled upon a good model of large 't Hooft coupling theories!

As a contrasting model, one could consider the digamma function,
$\psi(x)$, which is the logarithmic derivative of the gamma
function, \beq \psi(x) = \frac{\Gamma'(x)}{\Gamma(x)} = -\gamma +
\sum_{k=0}^{\infty} \left(\frac{1}{k+1} - \frac{1}{x+k}\right). \eeq
This function has the asymptotic expansion: \beq \psi(x+1) \approx
\log(x) + \frac{1}{2x} - \sum_{n=1}^{\infty} \frac{B_{2n}}{2n
x^{2n}}, \eeq so that it has the right sort of behavior to match
perturbative asymptotics. This function has been discussed as a model
of a QCD-like spectral function a number of times in the literature, beginning
with Ref. \cite{ShifmanQuarkHadronDuality}. It allows us to
build a model of the spectral
function with the first pole in the same location  (setting the
$\rho$ mass), $s = 1/2$, as: \beq \Pi_R(s) = \psi\left(s -
\frac{1}{2}\right). \eeq Note that this function, unlike the Pad\'e
result, has poles evenly spaced in $m^2$, not in $m$, so that it
matches expectations for QCD. We show this function in the plots for
comparison.

\begin{figure}[!h]
\begin{center}
\includegraphics[width=6.5cm]{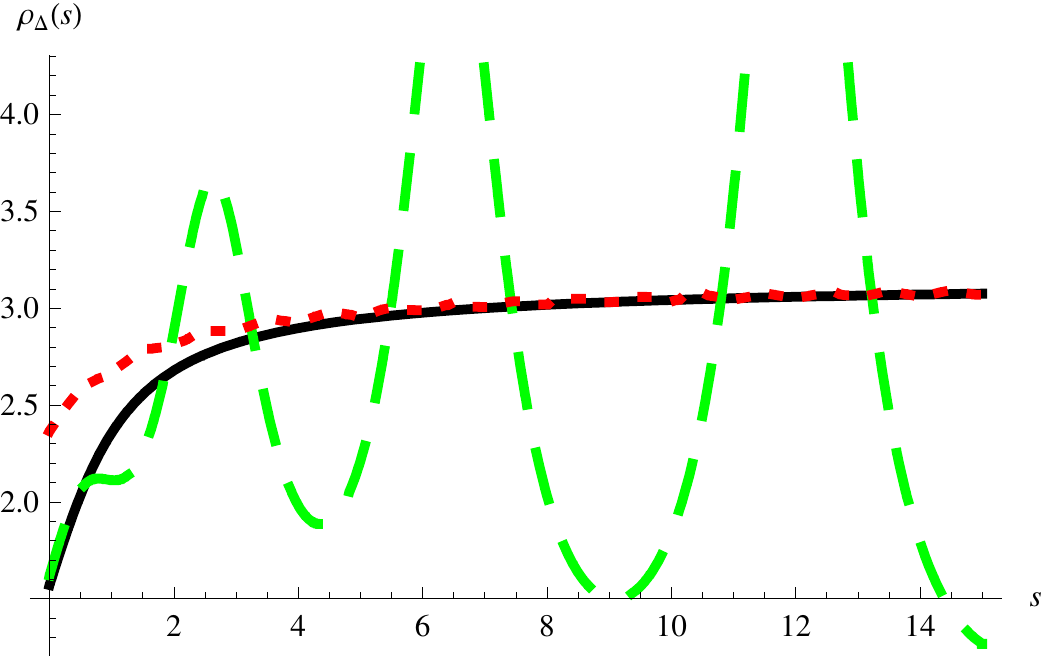}
\end{center}
\caption{Smeared spectral function, $\frac{1}{2 i}(\Pi(s + i) -
\Pi(s-i))$: black solid line, perturbation theory; green dashed
line, Pad\'e approximation; red dotted line, digamma function. The
digamma function with proper Regge physics provides a reasonably
good match, while the Pad\'e approximation oscillates about the
perturbative result with large amplitude.} \label{fig:rhosmear}
\end{figure}

\begin{figure}[!h]
\begin{center}
\includegraphics[width=6cm]{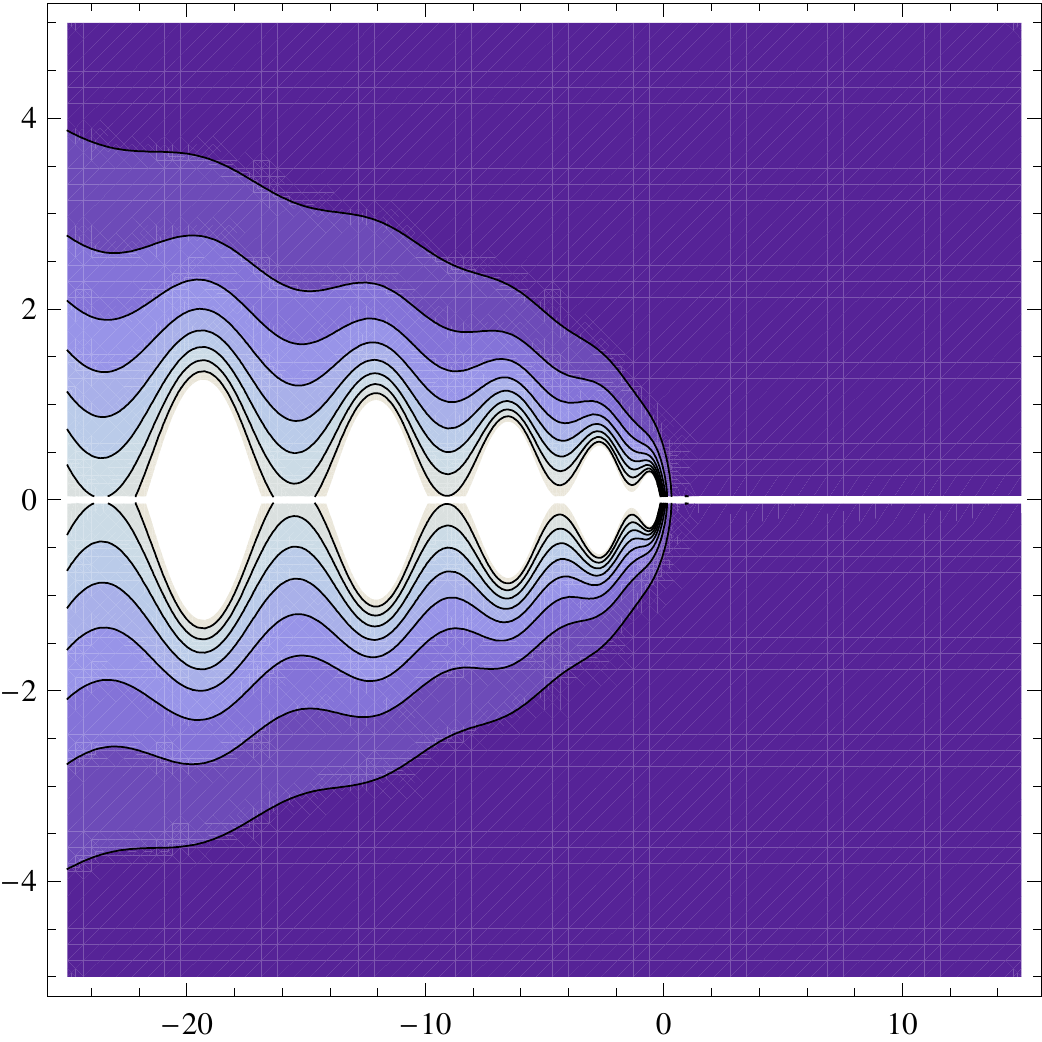}\includegraphics[width=6cm]{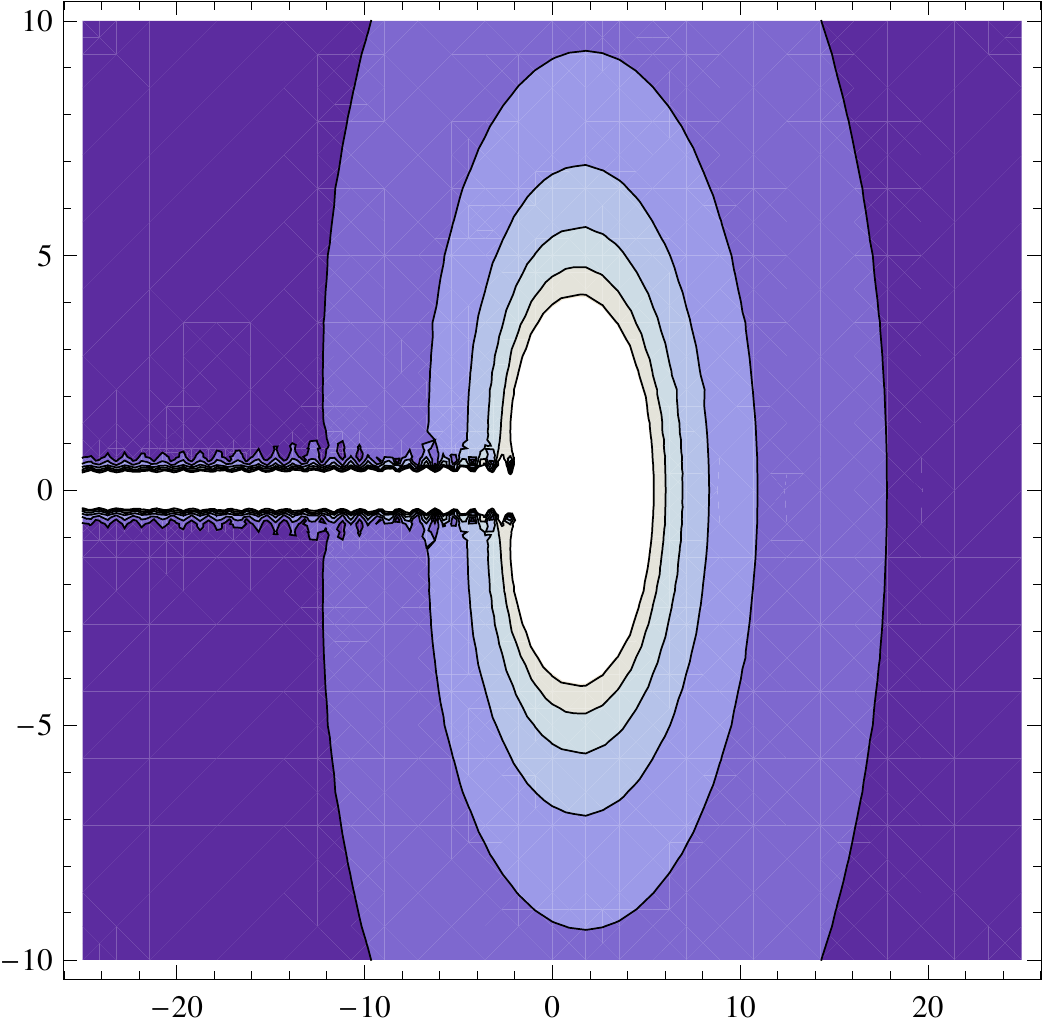}
\end{center}
\caption{Contours of constant fractional difference,
$\frac{\Pi(q^2)-\log(q^2)}{\log(q^2)}$, between the leading-order
perturbative result and the full two-point function. At left is the
Pad\'e case $\Pi_P$ and at right the digamma function $\Pi_R$. White
regions correspond to large fractional disagreements, and dark
regions to very good agreement. The digamma function disagrees with
PT in a circle of order $\Lambda_{QCD}$, and very close to the
Minkowski axis, while the Pad\'e approximant disagrees in a growing
wedge in the Minkowski region.} \label{fig:contours}
\end{figure}

The obvious question is which of these approximations one should
take more seriously for the case of large $N$ and small 't Hooft
coupling. On the face of it both of these functions reproduce the
perturbation theory results quite precisely. However, we will
argue that that despite the good agreement with PT for the deep
Euclidean region the Pad\'e approximation actually does not match PT
over the whole complex plane all that well, and that other functions
like $\Pi_R(s)$ exist that will capture the global features of PT
much better. To understand why the Pad\'e approximation fails we
need to understand the region of momenta where the Pad\'e
approximation is bad. Clearly it is doing extremely well for the
deep Euclidean region. However, some of the Minkowski region should
also be attainable through PT, at least when we are away from the
poles, i.e. away from the real axis $s>0$. One particularly simple
quantity of this sort is a smeared version of the spectral function
$\rho$:
\begin{equation}
\rho_{\Delta}(s) = \frac{1}{2i} \left(\Pi(s + i \Delta) - \Pi(s - i
\Delta)\right).
\end{equation}
A function of this sort has been discussed by Poggio, Quinn, and
Weinberg~\cite{PQW}. Many comparisons of experimental data on QCD
with PT implicitly rely on an approximate {\em
local} quark-hadron duality that allows a result smeared over
hadronic resonances to match PT. At low orders of
PT, the choice $\Delta \sim 1$ is effectively an
infrared cutoff at about the scale $\Lambda_{QCD}$. (In fact, since
$s = 1/2$ here corresponds to the $\rho$ mass squared, this acts
like an IR cutoff of about 1 GeV.) However, the Pad\'e approximant
disagrees quite sharply with the expected PT answer, as can be seen
in fig.~\ref{fig:rhosmear}. It would be rather surprising if such a
smeared perturbative calculation could disagree so strongly with the
true hadronic answer at large values of Minkowski momentum! On the
other hand, the smeared spectral function calculated from the
digamma function $\Pi_R$ will still be a very good approximation to
the PT answer even for Minkowski momenta. The agreement is shown
over the complex plane in Figure \ref{fig:contours}. Note that
$\Pi_R$ has larger error in a $\Lambda_{QCD}$-size region around the
origin, but otherwise disagrees only in a narrow strip along the
Minkowski axis. (Of course, one should not expect that the agreement
is good until $|Q^2| \gg \Lambda_{QCD}$, so this is entirely as expected!)
On the other hand, the Pad\'e result $\Pi_P$ has
large error in a wedge that grows progressively larger as one goes
deeper into the Minkowski region.

One would hope that one could complete Migdal's program of uniquely
fixing the appropriate meromorphic function relevant for large $N$
QCD by also requiring a good match to the Minkowski region. An
attempt in this direction is described in detail in Appendix A,
where we do show that the spacing between the poles
$m_{n+1}^2-m_n^2$ should not be growing, and that the simplest
meromorphic function satisfying all constraints from PT has
asymptotic properties agreeing with those of the digamma function.
However, we could not exclude some more exotic possibilities, so we
conclude that  Migdal's hope of finding a systematic procedure to
convert perturbative results to meromorphic functions does not seem
tractable, at least when only the two-point function is considered.

To summarize this section: we found that in order to apply Migdal's
program of analytically continuing the perturbative result one
should use the digamma function rather  than the Pad\'e
approximation. The use of such meromorphic functions with QCD-like
spectra has been previously discussed; see, e.g., Refs.
\cite{ShifmanQuarkHadronDuality, OPEtoSpectrum}. The result obtained by Pad\'e
approximation is a good guide to the large 't Hooft coupling
spectrum (that is the case when AdS is a good
description). Of course, at large 't Hooft coupling PT
breaks down, so it should not be surprising to see that there
can be a qualitatively different spectrum for a large 't Hooft
coupling. In the next section we will discuss a very simple toy
model that illustrates how the spectrum can change as the 't Hooft
coupling changes.

%%%%%%%%%%%%%%%%%%%%%%%%%%%%%%%%%%%%%%%%%%%%%%%%%%%%%
%%%%%%%%%%%%%%%%%%%%%%%%%%%%%%%%%%%%%%%%%%%%%%%%%%
\section{A toy model of bound states: where have all the KK modes gone?}
\label{sec:toyspectrum}
\setcounter{equation}{0} \setcounter{footnote}{0}
%%%%%%%%%%%%%%%%%%%%%%%%%%%%%%%%%%%%%%%%%%%%%%%%%%%%%
%%%%%%%%%%%%%%%%%%%%%%%%%%%%%%%%%%%%%%%%%%%%%%%%%%%%%

We have seen that a QCD-like spectrum is much more consistent with
perturbative calculations than a Pad\'e or AdS/QCD-like spectrum. It
would be interesting to understand in more detail how the spectrum
can change as one varies the 't Hooft coupling. In this section we
will discuss a very simple  toy quantum-mechanical model of bound
states in which the spectrum can change between QCD-like and
AdS/QCD-like. We make no claim that this is an accurate model of the
dynamics of any quantum field theory, but it does provide some
intuition about how strong coupling at short distances can change
the spectrum.

For studying the spectrum of mesons in a QCD-like theory, we expect
that we have a confinement potential well-approximated by the
Cornell form, \beq V(r) = -\frac{\lambda}{r} + \sigma r, \eeq with
$\sigma$ the string tension. The spectrum of mesons is, at least
qualitatively, determined by  a relativistic bound-state calculation
involving this sort of potential. (There have been various studies of
meson and glueball spectra in Coulomb gauge QCD,
which resemble these potential
models, including some which are quite sophisticated \cite{CoulombGauge}.)

What is notable about the  large $\lambda$ examples is that
the non-QCD-like states have  masses of order $M$ where $M\sim
\lambda^{-1/4} \sqrt{\sigma} \ll \sqrt{\sigma}$. This suggests that
these bound states involve physics at distances where the $\sigma r$
term of the potential is not yet dominant. Rather these states seem
to be related to the short-distance part of the potential, and so
they are likely better characterized as deeply bound mesons due to
the Coulomb nature of the potential at short distances. In QCD,
there is only one scale, $\Lambda_{QCD}$, and it's difficult to see
how such states could arise. So let's turn our attention away from
QCD and instead think about ${\cal N}=4$ super-Yang-Mills deformed
by mass terms to an ${\cal N}=1$ confining theory. The mass terms
give one scale, $m$, at which we expect strongly-bound mesons at
large $\lambda$, where the string tension is $\sqrt{\lambda} m^2$.

The static potential is computed along the lines discussed by
\cite{StaticPotentialAdSCFT}. In  particular for RS (as discussed in
\cite{BFBF}) it will be Coulomb-like until the string is long enough
to reach the IR brane; at that point, any additional string length
lies along the IR brane and the potential grows linearly. In other
words,
\begin{equation}
V(r) = \left\{ \begin{matrix}
-\frac{c}{r}, &~& r \leq r_c \\
-\frac{c}{r_c} + \sigma (r - r_c), &~& r \geq r_c
\end{matrix} \right.
\end{equation}
Here
\begin{equation}
c = \frac{4 \pi^2 \left(2 g_{YM}^2 N\right)^{1/2}}{\Gamma\left(\frac{1}{4}\right)^4} \approx 0.3231 \sqrt{\lambda},
\end{equation}
as in the AdS/CFT calculation. We can read off from Maldacena's
calculation of the static quark-antiquark  potential that the string
just touches the IR brane at $z_{IR}$ when the quark and antiquark
are separated by a distance
\begin{equation}
r_c = z_{IR} \frac{(2\pi)^{3/2}}{\Gamma\left(\frac{1}{4}\right)^2} \approx 1.198 z_{IR}.
\end{equation}
Finally, the string tension is given in terms of the 't Hooft coupling and the location of the IR brane:
\beq
\sigma = \frac{1}{2\pi\alpha'} \frac{R^2}{z_{IR}^2} = \frac{\sqrt{\lambda}}{2 \pi z_{IR}^2}.
\eeq

Our approach will be to use this potential to model the confining
force between a scalar quark and anti-quark in a simple toy model of
relativistic quantum mechanics. That is, we solve for radial
excitations that are eigenstates of the Hamiltonian \beq H =
\sqrt{p_r^2 + m^2} + V(r), \eeq where $p_r$ is the radial component
of the momentum. One can think of this as a very rough
approximation to a Hamiltonian formalism for the theory in Coulomb
gauge, approximating the Coulomb potential by the Wilson-loop
potential, although strictly speaking they are known to differ \cite{Zwanziger}.

\begin{figure}[!h]
\begin{center}
\includegraphics[width=6.5cm]{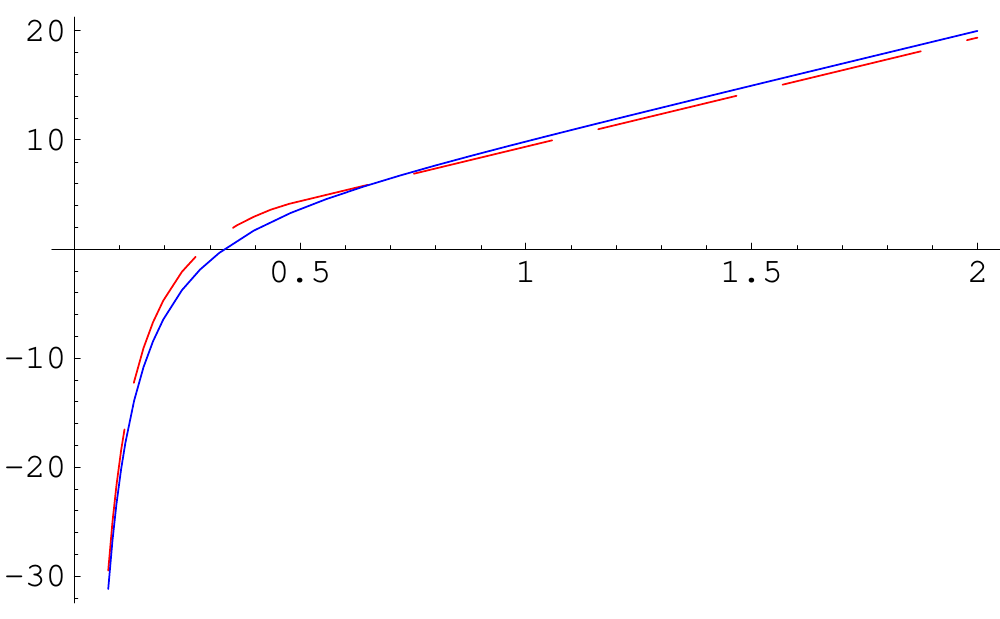}
\end{center}
\caption{Comparison of the RS potential (red, dashed), shifted by a constant value of 10.5, and our analytically useful replacement of it (blue, solid).}
\label{fig:vcomp}
\end{figure}

We will make one further approximation in order to do more of the
problem analytically, which facilitates an easy numerical solution
for the eigenstates. Namely, we will use a variational ansatz with
basis functions of the form $e^{-\beta r} L^2_n(2 \beta r)$, and we
would like a potential $V(r)$ that has analytically calculable
overlap with these basis functions. In particular we take the
parameters of the model to be specified by $m = 1$, $\sigma = 10$,
$c = 3$ (note that with these parameters $\sqrt{\lambda} \approx
9.28$ is large), and we approximate $V(r)$ by the more analytically
tractable: \beq V_{mod}(r) = -\frac{3}{r} \exp(-3 r) + 10 r, \eeq in
effect changing the sudden $\theta$-function like cutoff to a
smoother exponential shut-off of the Coulomb part of the potential.

This modified potential agrees quite well with the original RS
potential, shifted by a constant value of 10.5, as  illustrated in
Figure \ref{fig:vcomp}. Because there is no clear correct definition
of the zero of the potential, we will ignore the constant shift. (Of
course, if we look for evenly spaced masses, a constant shift in
energy eigenvalues is irrelevant; if we look for evenly spaced {\it
squared} masses, a constant shift will matter, but will become less
important at high enough excitation number.)

\begin{figure}[h]
\begin{center}
\includegraphics[width=6.5cm]{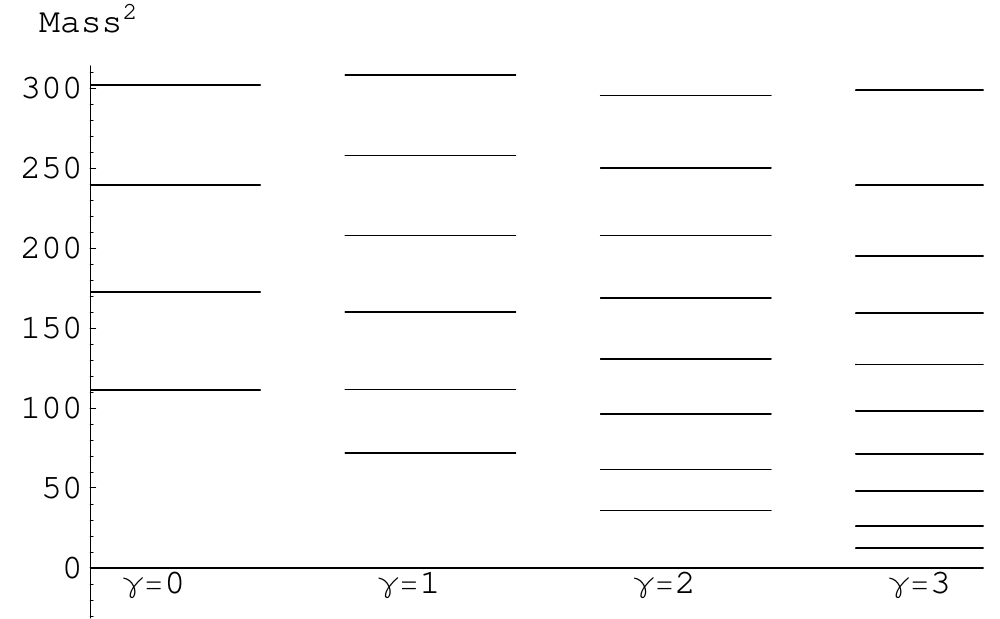}
\end{center}
\caption{Spectrum of our toy Hamiltonian .}
\label{fig:toyspectrum}
\end{figure}

Finally, to get a toy model of the transition from a QCD-like
spectrum to an RS-like one,  we vary the coefficient of the first
term in $V_{mod}(r)$: \beq V_{\gamma}(r) = -\frac{\gamma}{r} \exp(-3
r) + 10 r. \eeq At $\gamma = 0$ we have only a linear term,
resembling a QCD-like theory where Coulomb attraction plays very
little role in the pattern of bound states, while at $\gamma = 3$ we
have a theory which more closely resembles RS. The squares of the
eigenvalues of the Hamiltonian $\sqrt{p_r^2 + m^2} + V_{\gamma}(r)$
are plotted for $\gamma = 0,1,2,3$ in Figure \ref{fig:toyspectrum}.
One can see that as the strong Coulomb attraction is turned on, the
low eigenvalues become closely spaced and take on a qualitatively
new character (although they change continuously). These are deeply
bound states, sensitive to the strong short-distance Coulomb
attraction and relatively insensitive to the long-distance linear
confinement. They are the toy model analogs of KK modes. Higher
eigenvalues are spaced successively farther apart, eventually
merging into the original pattern of linearly spaced mass-squared.
They are the toy model analogs of resonances on linear Regge
trajectories. Note that at $\gamma = 0$ one finds roughly even
spacing consistent with $m^2(n+1) - m^2(n) = \frac{1}{\alpha'}$.

A more reliable and field-theoretic approach to modeling the effect
of strong short-distance  Coulomb attraction on bound states would
be to use Bethe-Salpeter equations, but already our toy model has
interesting physics that suggest a picture of how the transition
from QCD-like to RS-like dynamics can happen. As a final remark on
this topic, we should point out that literature beginning from a
``soft wall metric" background \cite{SoftMetric} computes AdS/QCD
potentials which are good matches to lattice data and which work
well phenomenologically in Salpeter calculations
\cite{AdSQCDPotential}. On the other hand, we will see in the next
section that event shapes in soft wall backgrounds are not QCD-like,
so if such potentials are to fit into a coherent framework for QCD
phenomenology, one would have to understand stringy corrections to
the effective action.

%%%%%%%%%%%%%%%%%%%%%%%%%%%%%%%%%%%%%%%%%%%%%%%%%%%%%%
%%%%%%%%%%%%%%%%%%%%%%%%%%%%%%%%%%%%%%%%%%%%%%%%%%%%%%
\section{Dynamics for large $N$, large $\lambda$: Spherical Event Shapes}
\label{sec:KKmodes}
\setcounter{equation}{0}
\setcounter{footnote}{0}
%%%%%%%%%%%%%%%%%%%%%%%%%%%%%%%%%%%%%%%%%%%%%%%%%%%%%%
%%%%%%%%%%%%%%%%%%%%%%%%%%%%%%%%%%%%%%%%%%%%%%%%%%%%%%

So far we have been discussing the two-point function of a current
with itself, which  determines the spectrum of associated resonances
and the total cross section of a process like $e^+ e^- \rightarrow$
hadrons. However, the real physics of $e^+ e^- \rightarrow$ hadrons
is much richer: in QCD, for instance, we know that such an event
will lead to a parton shower that produces collimated jets of
particles moving in a small number of directions. We now want to
turn our attention away from the crude questions of spectrum and
cross section and toward the richer questions about the structure of
events. This will mean going to finite $N$ and considering how
resonances decay and broaden into continuum structure.

Our tool for understanding whether an event is spherical or jetty
will be the ``sphericity" \cite{sphericity}, which is defined as $S
=\frac{3}{2} (\lambda_2 + \lambda_3)$ where $\lambda_2$ and
$\lambda_3$ are the two smallest eigenvalues of the matrix: \beq
S_{jk} = \frac{\sum_i p^{(i)}_j p^{(i)}_k}{\sum_i \left|{\vec
p}^{(i)}\right|^2}, \eeq where $p^{(i)}$ is the 3-momentum of the
$i^{\rm th}$ final-state particle. The sphericity is 1 for a
completely spherical event, and 0 for two back-to-back particles.
Perturbative QCD predicts small sphericities of order few$\times
0.01$.

Extra dimensions, in general, predict a sizable sphericity. To see why,
recall that in a flat periodic extra dimension, or a flat interval
with Neumann boundary conditions on both ends, there is a discrete
conserved 5D momentum, or ``KK number." The wave functions are
cosines and triple overlaps vanish except when the mass of the highest
state is precisely the sum of the masses of the two lower states. This
case is not very interesting; there is no phase space for decays. In a
UED model, the exact KK number conservation is broken by boundary
masses or by localized boundary kinetic terms. However, in this case, as in
many other theories with an extra dimension, there is still an
approximately conserved KK number. In particular, in a two-body
decay of a heavy KK mode, the masses of the two daughters are very close
to being equal to the mass of the parent; as a result, there is little
phase-space and the daughters are not very boosted. Thus, if
they subsequently decay, the new daughters will again not have
much of a boost, and the momenta of the various particles in the
final state will not be strongly correlated with any particular direction.
This leads to spherical events.

\begin{figure}[h]
  \centering
  \includegraphics[width = 0.25\textwidth]{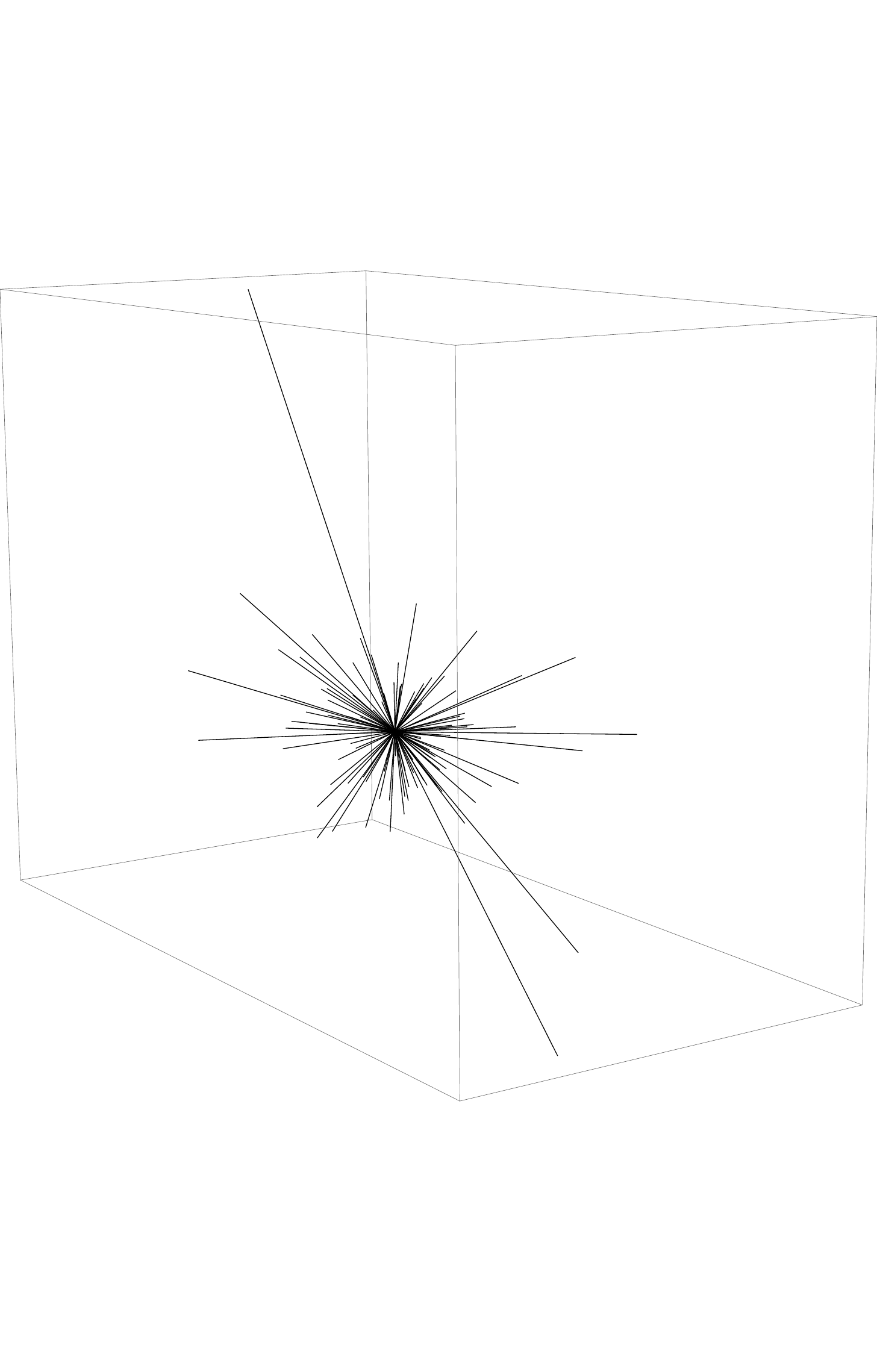}\hspace*{3cm} \includegraphics[width = 0.6\textwidth]{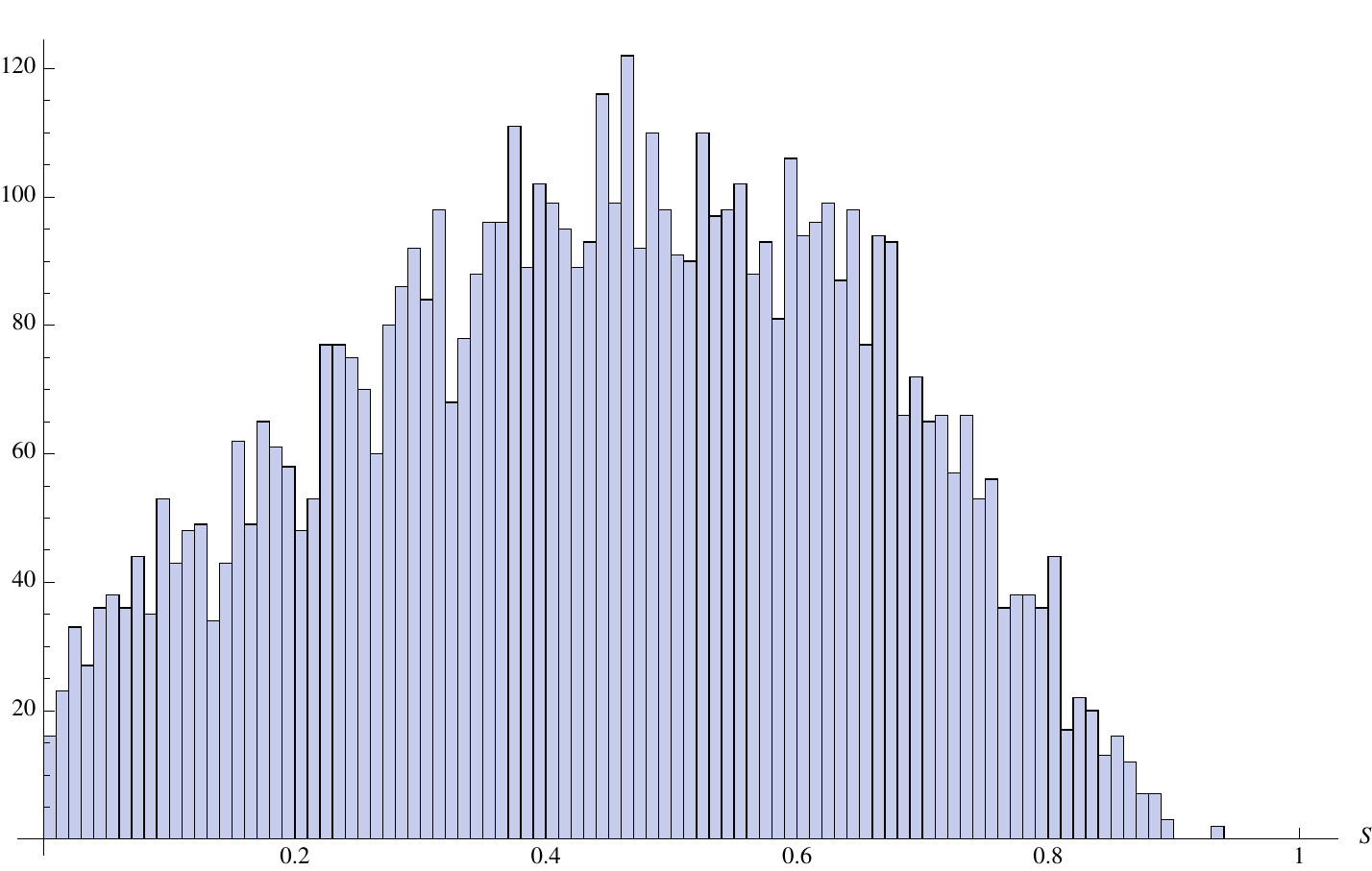}
  \caption{On the left: typical spherical event from the decay of the 200th gauge KK mode in flat
  space. On the right: the sphericity distribution from 6000 decays of the 200th gauge KK mode in flat space.
  }\label{fig:sphericalevent}
\end{figure}

In particular, Figure \ref{fig:sphericalevent} shows one example of
the shape of such an event. This is the decay of a heavy KK mode of
a gauge boson in flat space with a Neumann boundary condition on one
end of the space and a Dirichlet boundary condition on the other.
The spikes radiated out represent the momenta of stable daughters at
the end of the decay chain, with length proportional to energy. Note
that daughters are flying out in all directions, and if we were to
place a spherical detector of large radius around the origin, it
would see a fairly uniform distribution of energy. In particular,
the sphericity is high. This is not a jetty event, but one closer to
the uniform energy distribution found in the conformal case
\cite{HofmanMaldacena}. The distribution of sphericity for such
decays is also shown in Figure \ref{fig:sphericalevent}, where we
see that an overwhelming fraction of events have fairly large
sphericities $>0.1$.

Although approximate conservation of KK number is most obvious
in flat space, it turns out that even in warped space it is still true.
As a first example we consider a scalar field
in an RS background with  cubic interaction: \beq S = \int d^4
x~\int_{z_{UV}}^{z_{IR}} dz~\left\{\left(\frac{R}{z}\right)^3
\left(\eta^{\mu \nu} \partial_\mu \phi \partial_\nu \phi -
\partial_z \phi \partial_z \phi\right) + \left(\frac{R}{z}\right)^5
g_5 \phi^3 \right\}, \eeq with $R$ the AdS curvature radius. We take
$z_{UV} \rightarrow 0$, requiring that wave functions be
normalizable at the boundary, and we impose a finite $z_{IR}$ where
Dirichlet boundary conditions are imposed. This means that our wave
functions are: \beq \psi_n(z) = \frac{z^2 J_2(m_n z)}{\sqrt{-R^3
z_{IR}^2 J_1(m_n z_{IR}) J_3(m_n z_{IR})/2}}, \eeq where $J_2(m_n
z_{IR})$ = 0 and the denominator serves to normalize the wave
function.

\begin{figure}[!h]
  \centering
  \includegraphics[width = 0.3\textwidth]{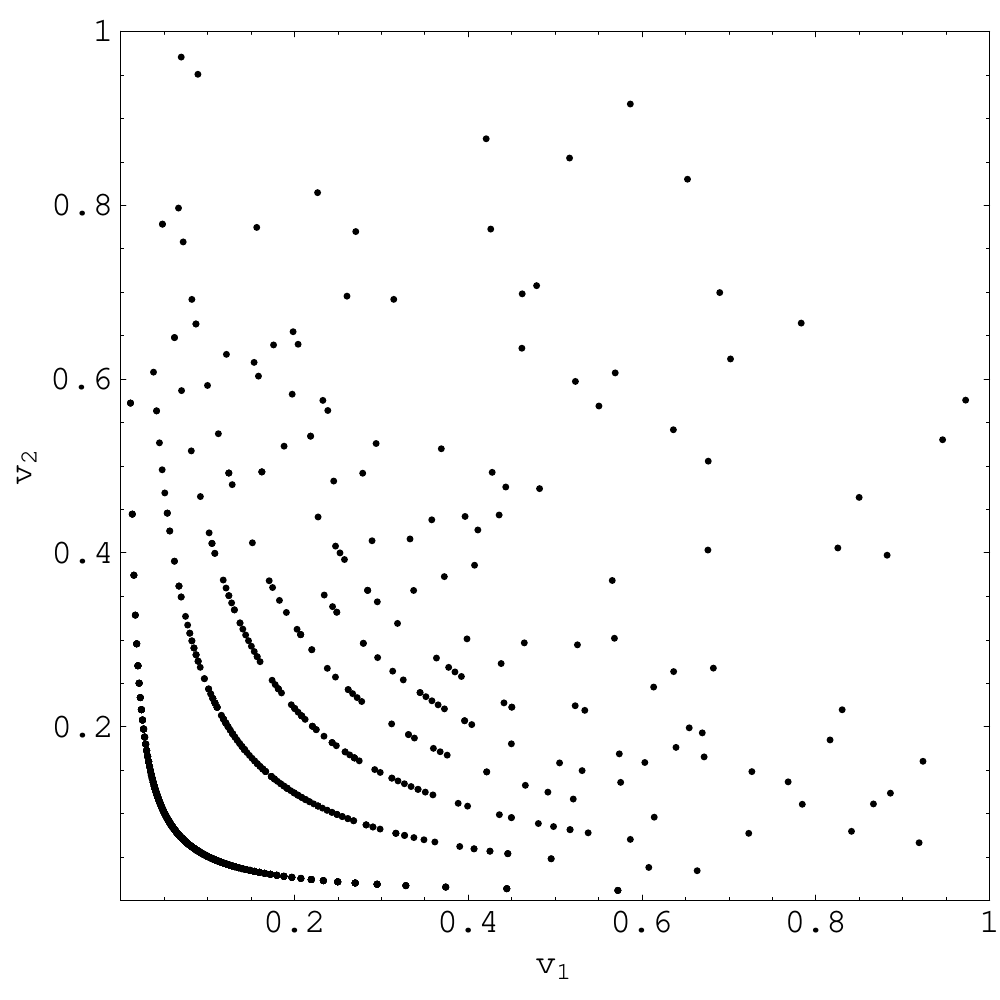}
  \caption{Distribution of the speeds $v_1$ and $v_2$ of the two Kaluza-Klein modes
  originating from the decay of the 100th scalar KK mode in RS. Note that typically the decay is to slow-moving particles.}\label{fig:kkspeeds}
\end{figure}

\begin{figure}[!h]
  \centering
  \includegraphics[width = 0.25\textwidth]{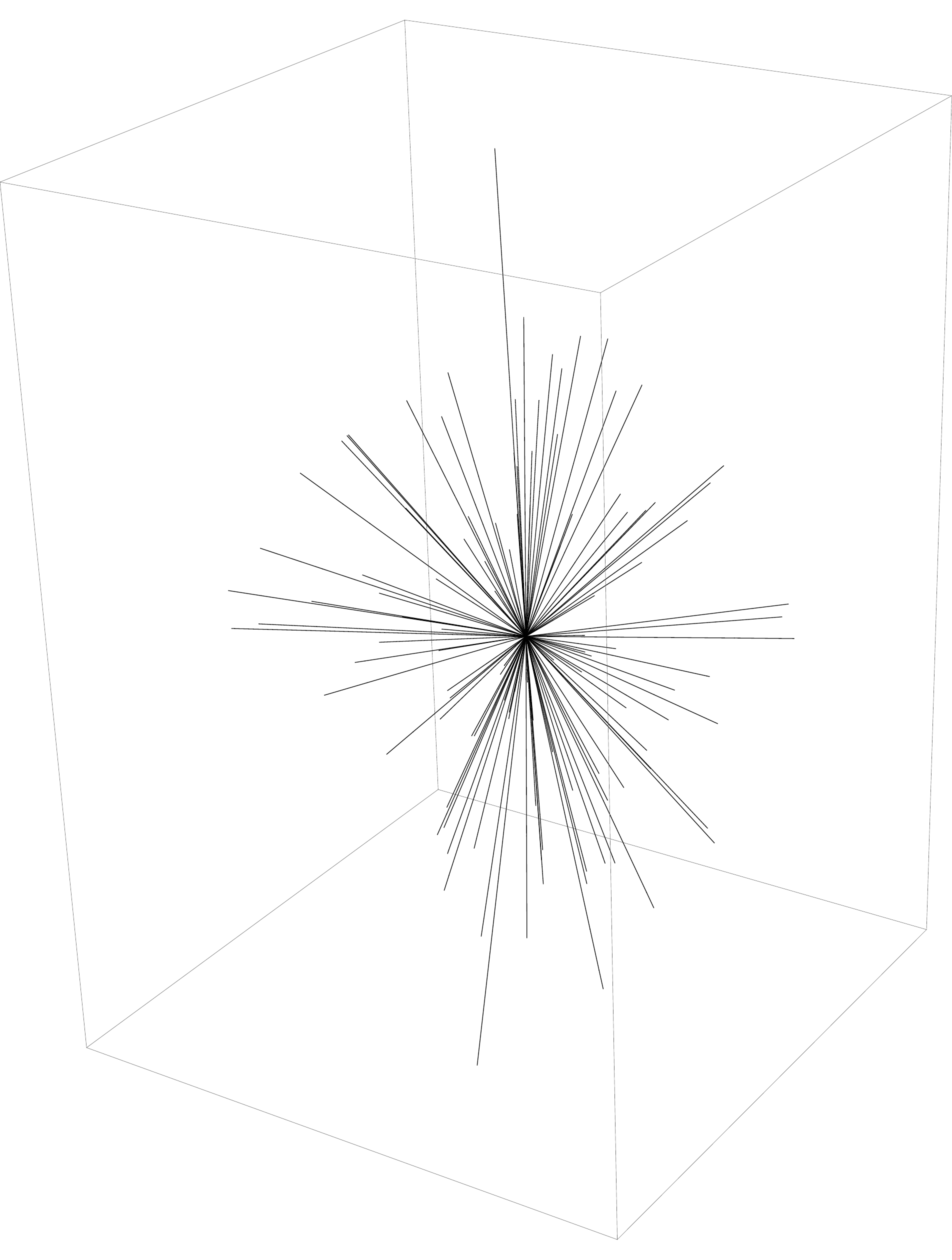} \hspace*{2cm} \includegraphics[width = 0.6\textwidth]{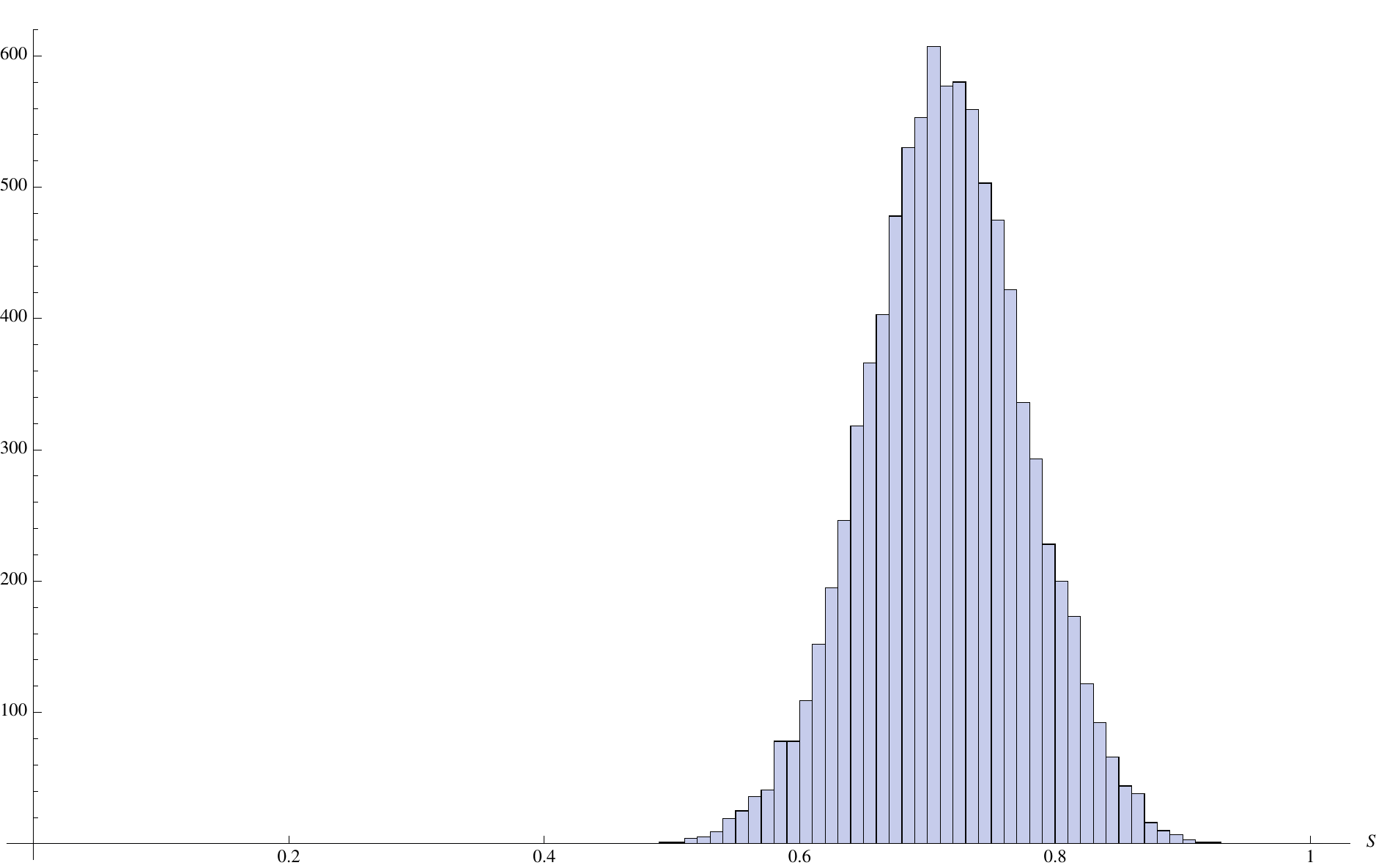}
  \caption{Left: a typical spherical event from the decay of the 200th KK gauge KK mode in AdS. Right: Sphericity distribution event from 9000
  decays of the 200th gauge KK mode in AdS space.}\label{fig:sphericaleventAdS}
\end{figure}

\begin{figure}[!h]
  \centering
  \includegraphics[width = 0.4\textwidth]{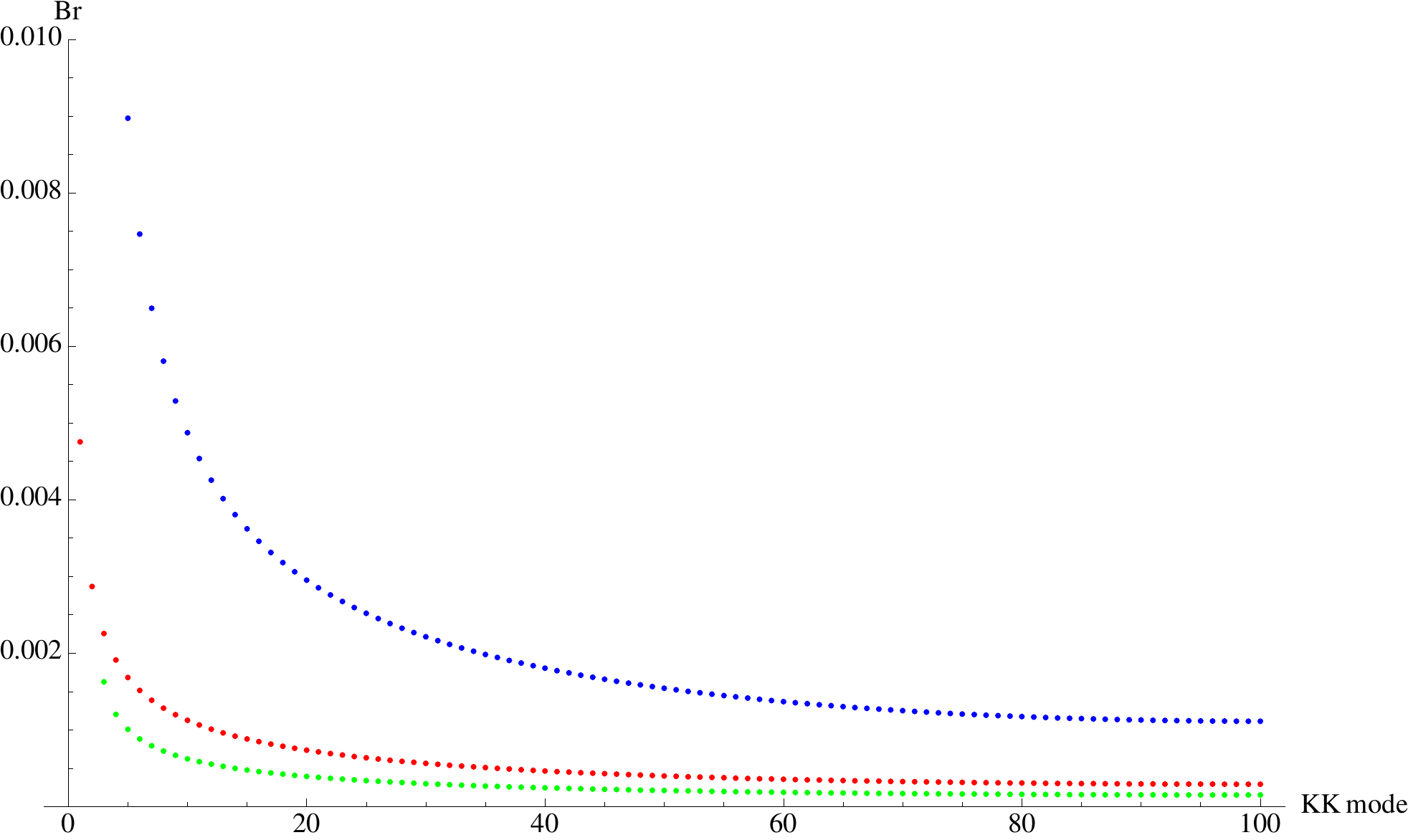}
  \caption{Branching fractions for the 200th KK mode with the KK number of one of the daughters on the horizontal axis.  In the top line (Blue) the KK numbers of the daughters add up to 198, in the second line (Red) they add to 199, in the third line (Green) they add to 197. }\label{fig:AdSBF}
\end{figure}

From these wave functions, we can compute couplings as
triple overlaps and then compute decay chains. While it is not
obvious from inspecting the wave function, it turns out that
the triple overlaps still show a strong tendency to almost conserve
KK number, so that decays with little phase space are preferred.
As an illustration of this, we plot in Figure \ref{fig:kkspeeds} the
distribution of the speeds, $v_{1,2} = |{\vec p}_{1,2}|/E_{1,2}$,
of the two KK modes originating from the decay of the 100th
scalar KK mode in RS. Notice that in almost every case, the decay
is to two relatively slow-moving, unboosted particles; the masses
of the two daughters sum to nearly the mass of the parent.
Just as in flat space, the tendency to prefer decays with little
phase space will lead to very spherical events. A typical such event
for the decay of a heavy gauge KK mode in RS is shown on the left
panel in Figure \ref{fig:sphericaleventAdS}, and the sphericity
distribution for many such events is shown in the right panel in
Figure \ref{fig:sphericaleventAdS}. Because the zeroes of Bessel
functions are very nearly linearly spaced, it turns out that the
number of the KK mode is almost conserved. This is illustrated in
Figure \ref{fig:AdSBF}. The 200th KK mode decays most often to two
modes with mode number summing to 198, then to 199, then to 197, and
so on.

RS backgrounds are similar to flat space, to the extent that they
predict mass spectra with nearly linear spacing. What happens if we
tune an extra-dimensional theory to achieve the Regge spacing $m_n^2
\sim n$, which holds in small 't Hooft coupling theories as
discussed in the previous sections? One might hope that
almost-conservation of KK number is lost and event shapes are very
different. We will now see that is not the case; despite the very
different spectrum, decays with little phase space are still
preferred and sphericities are large. Our starting point will be a
gauge field in the by now familiar ``soft-wall" model of AdS/QCD, in
which the metric is AdS but the action is~\cite{softwall}: \beq
S & = & \int d^5 x~e^{-\Phi}\sqrt{g} \frac{-1}{4 g_5^2} F_{MN} F^{MN}, \\
e^{-\Phi} & = & e^{-\Lambda^2 z^2}.
\eeq
The exponentially damped dilaton leads to an equation of motion
that can be rewritten as a Schr\"odinger equation for the eigenvalues
$m_n^2$ with harmonic oscillator potential $\sim z^2$ at
large $z$, which yields the desired spectrum. In the original
soft-wall model, the dilaton is simply assumed to have this profile,
due to some unknown dynamics.

In this model, triple overlaps of wave functions vanish;
we must deform it slightly to get a model in which we can calculate
decays. As a simple deformation, we include a ``UV brane" at
$z = z_{UV}$ and impose Dirichlet boundary conditions there (the other
boundary condition is simply normalizability at $z \rightarrow \infty$).
This gives a small deformation to the wave functions and masses
and allows us to compute decay chains. It turns out that the correct
wave functions are given by:
\beq
\psi_n(z) = c_n z^2 U\left(\frac{-m_n^2}{4 \Lambda^2} + 1, 2, z^2\right),
\eeq
with $U$ a confluent hypergeometric function. The condition
$U\left(\frac{-m_n^2}{4 \Lambda^2} + 1, 2, z_{UV}^2\right) = 0$
determines the masses $m_n$ and the constant $c_n$ is chosen
to give canonically normalized KK modes.

\begin{figure}[!h]
  \centering
  \includegraphics[width = 0.3\textwidth]{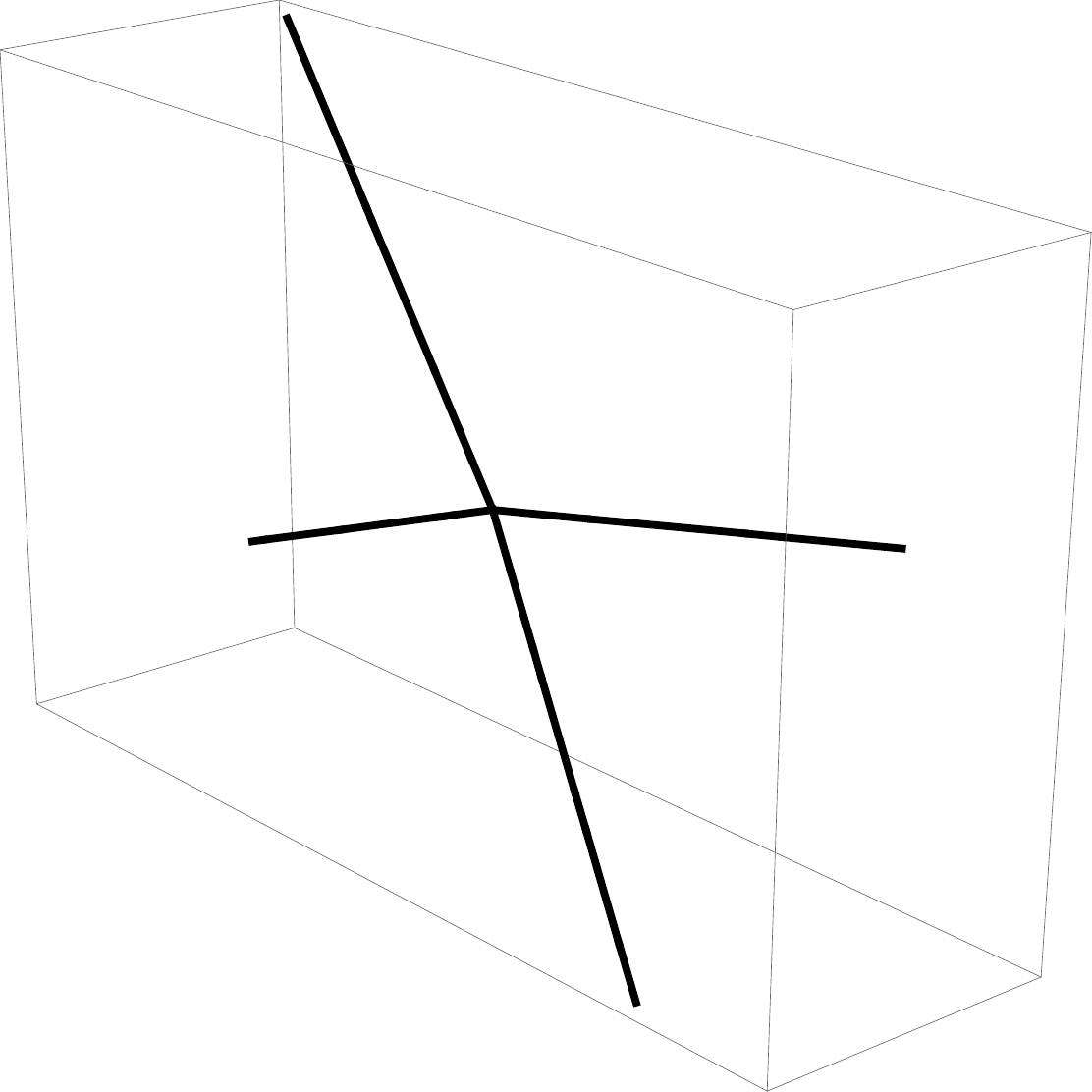}  \hspace*{2cm} \includegraphics[width = 0.3\textwidth]{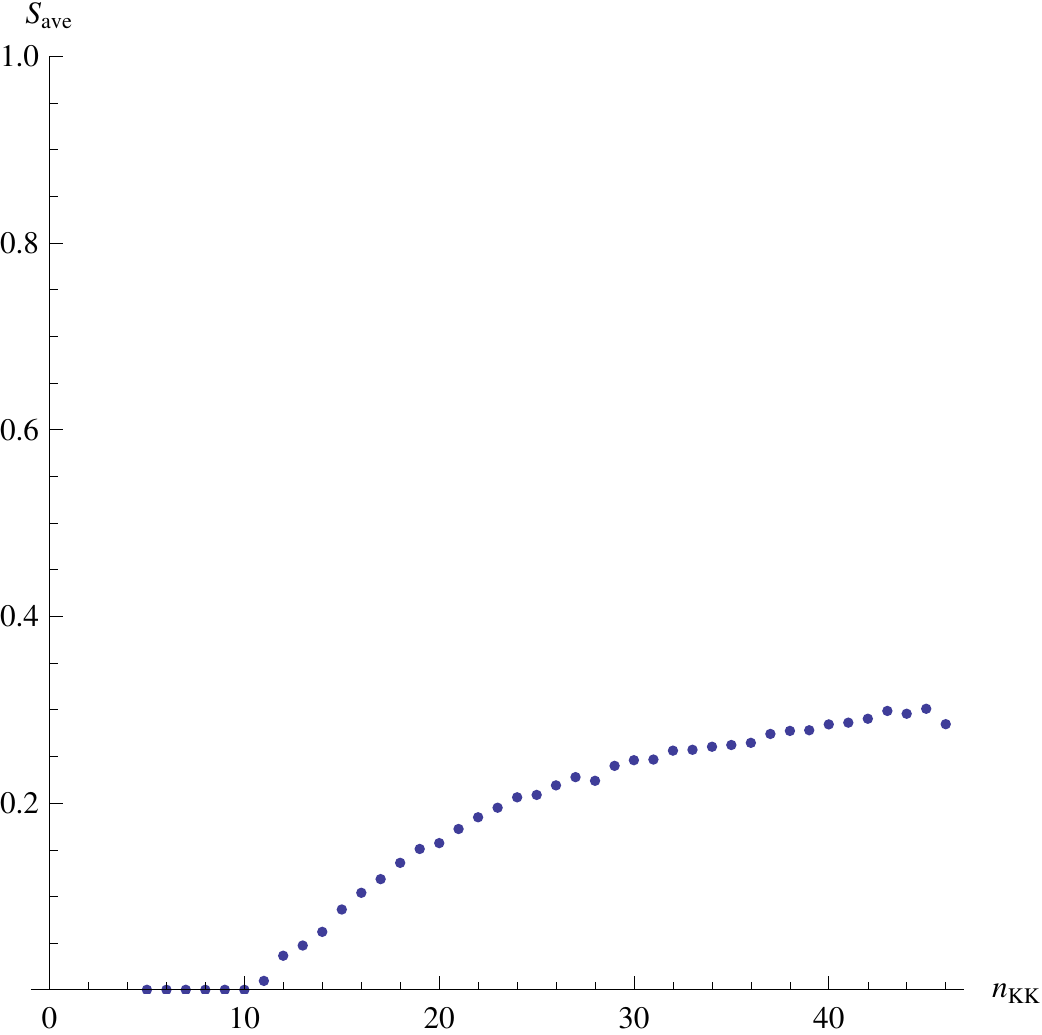}
  \caption{At left: a typical spherical event from the decay of a heavy (the 46th mode) soft-wall gauge boson.
  The UV brane was chosen to be at $z_{UV}=0.01 /\Lambda_{QCD}$.
  At right: average sphericity in 500 decays of the n$^{th}$ KK gauge boson in the soft-wall
  background. The horizontal axis is the KK number and the vertical the average spehricity of 500 decays of that mode.}
  \label{fig:sphericaleventsoftwall}
\end{figure}

In Figure \ref{fig:sphericaleventsoftwall} we show a typical
spherical event, and the distribution of sphericities. The numerical
treatment of the hypergeometric $U$-function is much harder than
that of Bessel functions, and due to the smaller spacing among the
masses there are fewer kinematically allowed decays for a given KK
mode. Nevertheless, even with a relatively low number of events the
trend toward final states with larger sphericities is quite clear.
This is again because despite the very different mass spectrum, the
wavefunction overlaps prefer the decays with small phase space and
correspondingly small boosts. The soft wall predicts more spherical
events which are un-QCD-like, despite its QCD-like mass spectrum.
Attempts to put the soft-wall background on a dynamical footing, as
in Ref. \cite{moresoftwall}, seem unlikely to remedy this problem
unless they involve inherently stringy dynamics. One perhaps more
promising direction, begun in Ref. \cite{KatzOkui}, is to begin with
the solvable large-$N$ 't Hooft model in 1+1 dimensions and attempt
to reformulate it as a 2+1 dimensional theory. Such an approach
could also clarify the relationship between holographic
wavefunctions and the internal structure of hadrons
\cite{BrodskyDeTeramond}. It remains to be seen how far this program
can be pushed, and whether it will lead to anything resembling the
soft wall. Also, note that demanding ``asymptotic freedom" in the
sense of a 5D dilaton which approaches a weak coupling value will
not change these event shapes. To truly model the dual of a weak
coupling theory, one would have to include light fields (string
modes) dual to the many operators with small anomalous dimension.
For now, it seems that there is still a major obstacle to obtaining
QCD-like dynamics from a 5D theory, one which will likely require
new techniques in string theory to be overcome. For now, we will
turn back to 4 dimensions, and look at how much older models of
hadrons generate event shapes that are much closer to those of QCD.

%%%%%%%%%%%%%%%%%%%%%%%%%%%%%%%%%%%%%%%%%%%%%%%%%%%%%%
%%%%%%%%%%%%%%%%%%%%%%%%%%%%%%%%%%%%%%%%%%%%%%%%%%%%%%
\section{Large $N$, Small `t Hooft Coupling: Flux tubes}
\label{sec:fluxtubes}
\setcounter{equation}{0}
\setcounter{footnote}{0}
%%%%%%%%%%%%%%%%%%%%%%%%%%%%%%%%%%%%%%%%%%%%%%%%%%%%%%
%%%%%%%%%%%%%%%%%%%%%%%%%%%%%%%%%%%%%%%%%%%%%%%%%%%%%%

\subsection{Flux tube Breaking, Directionality, and Jets}

If, as we have argued, the mesons of RS do not smoothly match onto
the mesons of a  QCD-like theory, what sort of mesons do? There is a
long history of modeling the highly excited states of QCD-like
theories using flux tubes. Among the reasons for doing this are the
idea that large $N$ QCD is a string theory because it has a
topological expansion \cite{'tHooft:1973jz}, the experimental fact
of Regge trajectories \cite{OldRegge}, the emergence of flux tubes on
the lattice \cite{LatticeStrings}, and general arguments about the behavior of
't Hooft versus Wilson loops \cite{'tHooft:1977hy,confinement}. Unlike in
theories with large 't Hooft coupling and good gravity duals, the
stringy physics in confining theories at small 't Hooft coupling
begins at the scale of the lightest resonances; there is no large
$\sqrt{\lambda}$ to provide a parametric separation of scales.

\begin{figure}[h]
  \centering
\includegraphics[width = 0.28\textwidth]{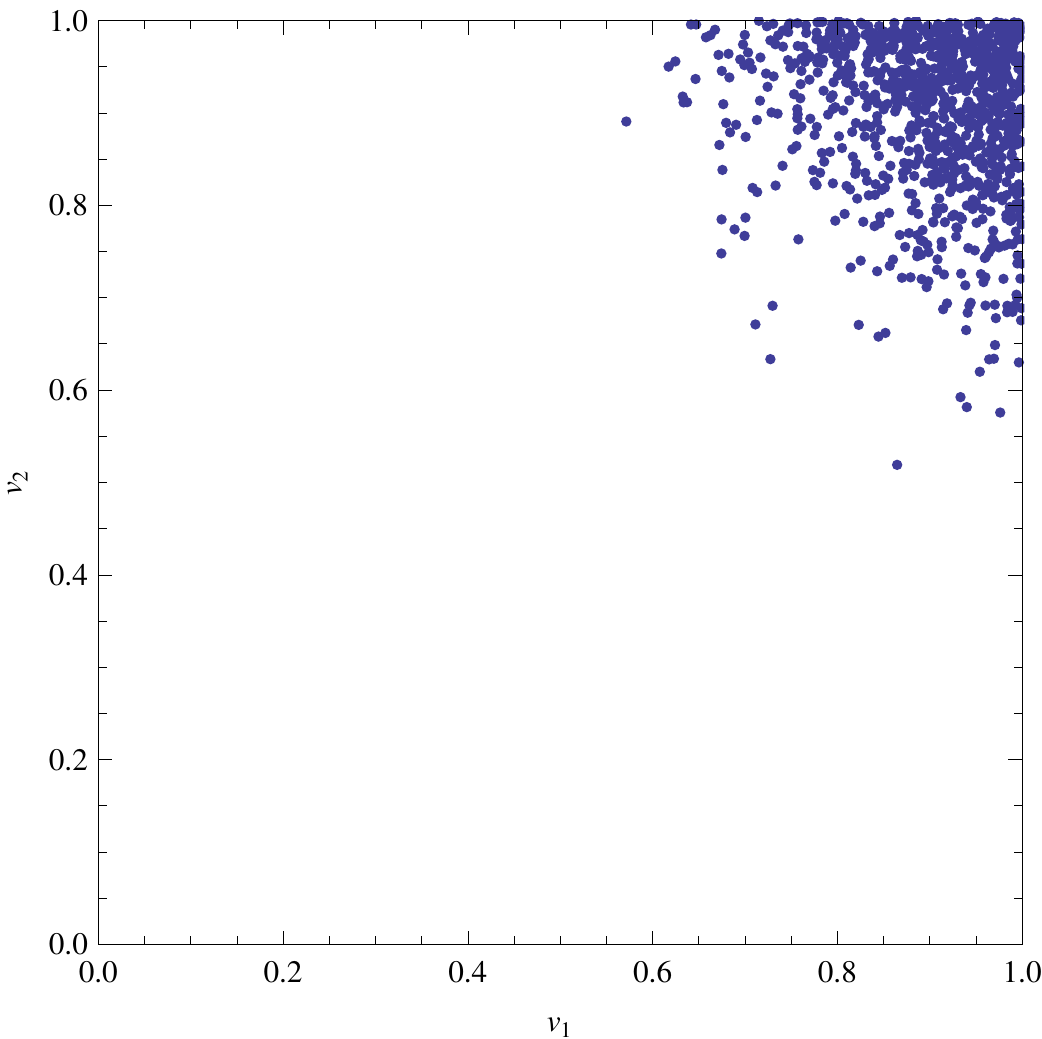}\includegraphics[width = 0.28\textwidth]{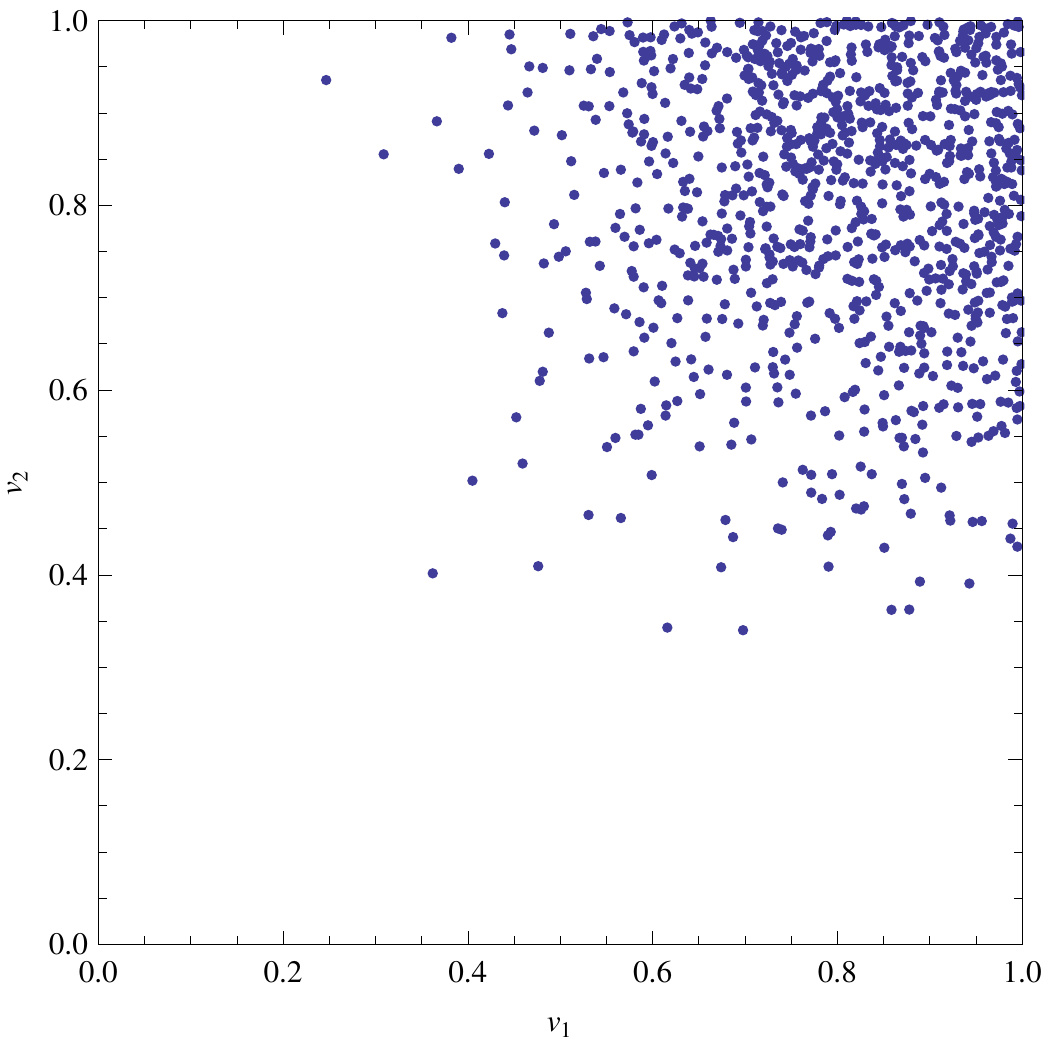}\includegraphics[width = 0.28\textwidth]{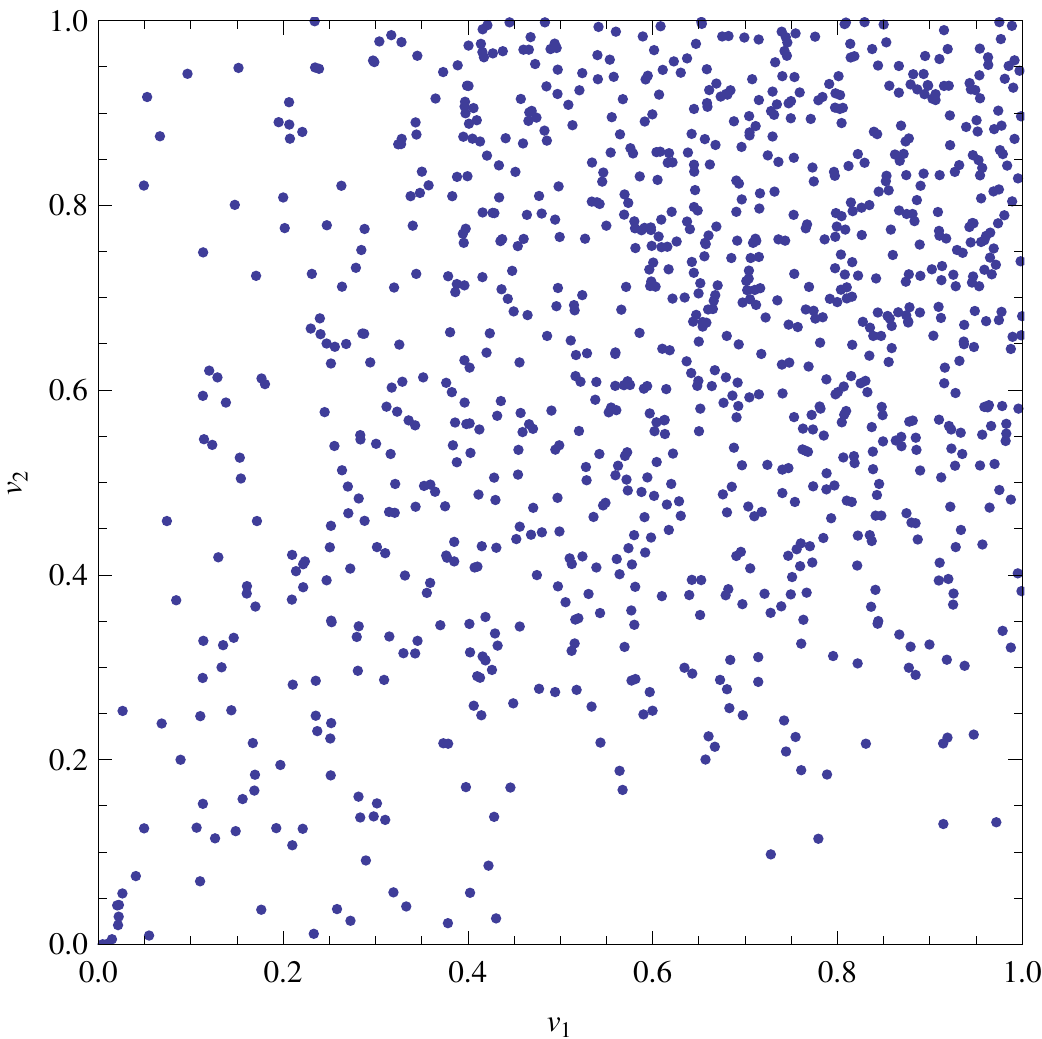}
  \caption{Distribution of the speeds $v_1$ and $v_2$ of the two flux tubes originating from breaking of an initial flux tube. The state begins with back-to-back quark and antiquark of 15 GeV momentum, and the string tension is 0.177 GeV$^2$. The decay rate per unit length is, from left to right, 0.01 GeV$^2$, 0.0025 GeV$^2$, and 0.000625 GeV$^2$. This corresponds to 99\% of the strings having decayed by the time they are one-quarter, half, and fully extended, respectively. Note that even in the latter case many decays are to fast particles.}\label{fig:ftspeeds}
\end{figure}

Before we look at more detailed modeling of flux tubes, let's begin by
thinking about the first breaking of the flux tube that would be produced, for example,  in an $e^+e^-$ collider. We begin with a quark and antiquark
moving in opposite directions, each with momentum $p$, and assume
they feel a linear potential, i.e. a constant attractive force $\sigma$ toward each
other. We assume the probability of breaking is uniform along the string
length; thus, as the string grows linearly, initially the decay probability
per unit time goes as $t^2$. If the string breaks immediately, then there
are massless quarks moving in opposite directions, and we have a
clean two-jet event, just as in PT. If the string breaks
when fully extended, we have two heavy static flux tubes of mass
$2px$ and $2p(1-x)$, with $x$ the fractional distance along the string
where the break occurs. These flux tubes will subsequently decay,
but the event will not be jetty. After the time $\tau_{osc} = p/\sigma$ at which
the string is fully extended, the quark and antiquark turn around and again
pick up speed, so at longer decay times there can still be many
events at which the decay products are boosted.

In Figure \ref{fig:ftspeeds} we show the distribution of speeds of the decay
products as the decay rate decreases. Provided the string decays
in a time smaller than that needed for the string to fully extend,
the first breaking is to high-speed objects, and we expect that their
further decays will widen the jets but maintain the qualitative ``jettiness"
of the event. The thrust axis of the jets will be the direction along which
the flux tube was stretched, established by the initial momentum of
the quark and antiquark, just as in PT.
Note that in order for the string to decay before the quark and
antiquark  reverse direction, we need $\tau_{dec} < \tau_{osc}$. The
decay time will be $\tau_{dec} \sim \frac{N}{p}$, so this tells us
that we need $N$  less than about $\frac{p^2}{\sigma}$. This is
consistent with large $N$ provided we look at strings that are
energetic enough.

If the decay rate is sufficiently small, so that $\tau_{dec} > \tau_{osc}$,
then we have slow daughters much more often, although the distribution
is still very different than that for KK modes in Figure \ref{fig:kkspeeds}.
Note, however, that for such small decay rates, the discreteness
of the large $N$ spectrum invalidates our way of thinking about
the decay. Given $m_n^2 \approx n \Lambda^2$, we have
\beq
m_{n+1} - m_{n} \approx \frac{\Lambda}{2 \sqrt{n}} \sim \frac{\sigma}{m_n}.
\eeq
Thus the spacing between subsequent modes is of order the inverse
of the oscillation time of our semiclassical string; if $\tau_{dec} > \tau_{osc}$,
the width is smaller than the interval between subsequent modes,
and it no longer makes sense to assume a continuous flux tube
that can break at any location. In this situation a quantum mechanical treatment is required, and the resulting  discreteness of the spectrum will play a role
in calculating the decays. We are unable to treat such long decay times  in the context of
this semi-classical model.

\subsection{A Simple Model of Flux Tubes}\label{sec:model1}

We have simulated a very simple toy model of the decay of an excited
flux tube, in  order to plot some event-shape distributions. The
literature has many more elaborate models, with varying degrees of
input from PT. Some early references are
\cite{Casher:1978wy,Gurvich:1979nq, Isgur:1983wj, Andersson:1983ia};
more recently, the ``gluon chain model" \cite{GluonChain} has
provided an example of a model that reflects both partonic and
flux tube limits of the theory, and is consistent with important
nonperturbative features of Yang-Mills theory like Casimir scaling
at intermediate distances and $N$-ality scaling at long distances.
(For a recent review of confinement in QCD-like theories, with
an emphasis on the role of these features, see Ref.~\cite{Greensite:2003bk}.)

Our model is a semiclassical one, based on linear confinement and
decay by  Schwinger pair production, following
Ref.~\cite{Casher:1978wy} as amended by Ref.
\cite{Glendenning:1983qq} so that energy is conserved when the flux
tube breaks. We begin with a quark and antiquark moving at high
momentum in opposite directions along the $z$ axis, separated by a
constant chromoelectric field filling a cylinder of radius $R$, so
that they are attracted by a constant force $\sigma$ (the string
tension). The total energy of the system is (taking the quarks to be
massless): \beq \label{eq:fluxham} E = \sum_{q} \left| \vec{p}_q \right| +
\sum_{\bar{q}} \left| \vec{p}_{\bar{q}} \right| + \sum_{q \bar{q}}
\sigma \left| \vec{x}_q - \vec{x}_{\bar{q}} \right|, \eeq where the
last term is a sum of the energy from string tension over
quark--antiquark pairs. We do not reconnect pairs, so that each
quark is associated with precisely one antiquark partner at all
times (changing only when there is a decay). The flux tube can decay
by the production of a quark-antiquark pair at some point along the
axis between quark and antiquark. The newly created pair will have
transverse momentum $p_T$ (orthogonal to the flux tube axis) and the
quark and antiquark will be separated by a
distance $d(p_T) = 2 E_T/\sigma$. The stretch
of string that has been annihilated between the two quarks then
compensates for the energy added by their initial momentum.
Transverse momenta are distributed according to a Gaussian, $dN(p_T)
\sim d^2 p_T~\exp\left(-\frac{\pi p_T^2}{\sigma}\right)$. Our
simulation takes small time steps, alternating a classical evolution
step and a random decay step. In the classical evolution step,
quark-antiquark pairs evolve according to their current momenta and
the color force between them, treating Eq. (\ref{eq:fluxham}) as
a Hamiltonian governing the evolution. (We never reconnect pairs, so each
quark remains connected to the same antiquark until the string
fragment splits, generating two shorter strings.) The random decay
step decays each string fragment with probability proportional to
the time step and to the length of the string fragment; decays
happen with $p_T$ (orthogonal to the quark-antiquark axis) chosen
from the Gaussian distribution and location of the decay chosen from
a uniform distribution on the string consistent with the newly
produced pair having the correct separation $d(p_T)$ and remaining
between the original quark and antiquark. This is a highly
simplified model, neglecting for instance any internal dynamics of
the string, but we expect it to get the grossest qualitative
features right, at a level that allows an approximate comparison
with the RS results and with QCD. It conserves energy (because the
newly created pair are always separated by an appropriate distance
along the string axis) and momentum (because the newly created pair
have equal and opposite momentum), but is not fully relativistically
invariant\footnote{In Appendix B we show that for large enough center of mass energies it does give a Lorent invariant description.}. In Section \ref{sec:model2} we will discuss the relativistic
string action and show that the first few decays computed from
it are qualitatively consistent with the results of this simpler model.

\begin{figure}[h]
  \centering
  \includegraphics[width = 0.4\textwidth]{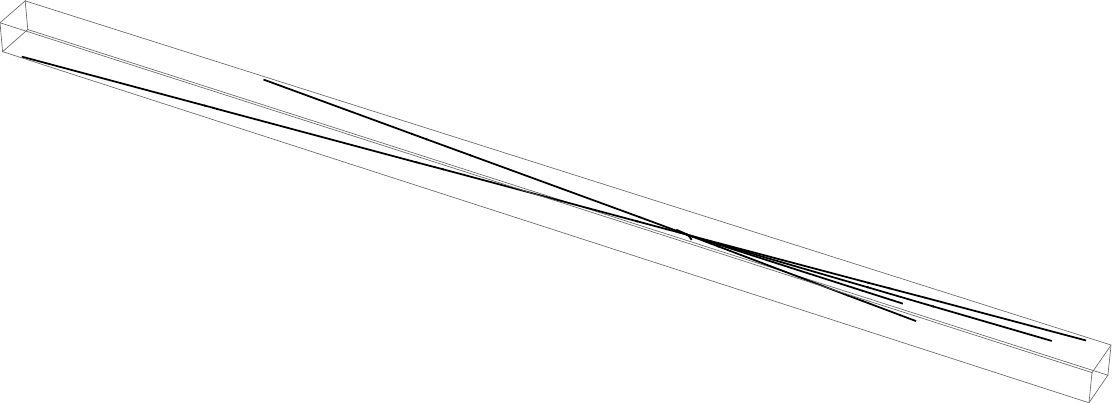}\hspace*{2cm} \includegraphics[width = 0.4\textwidth]{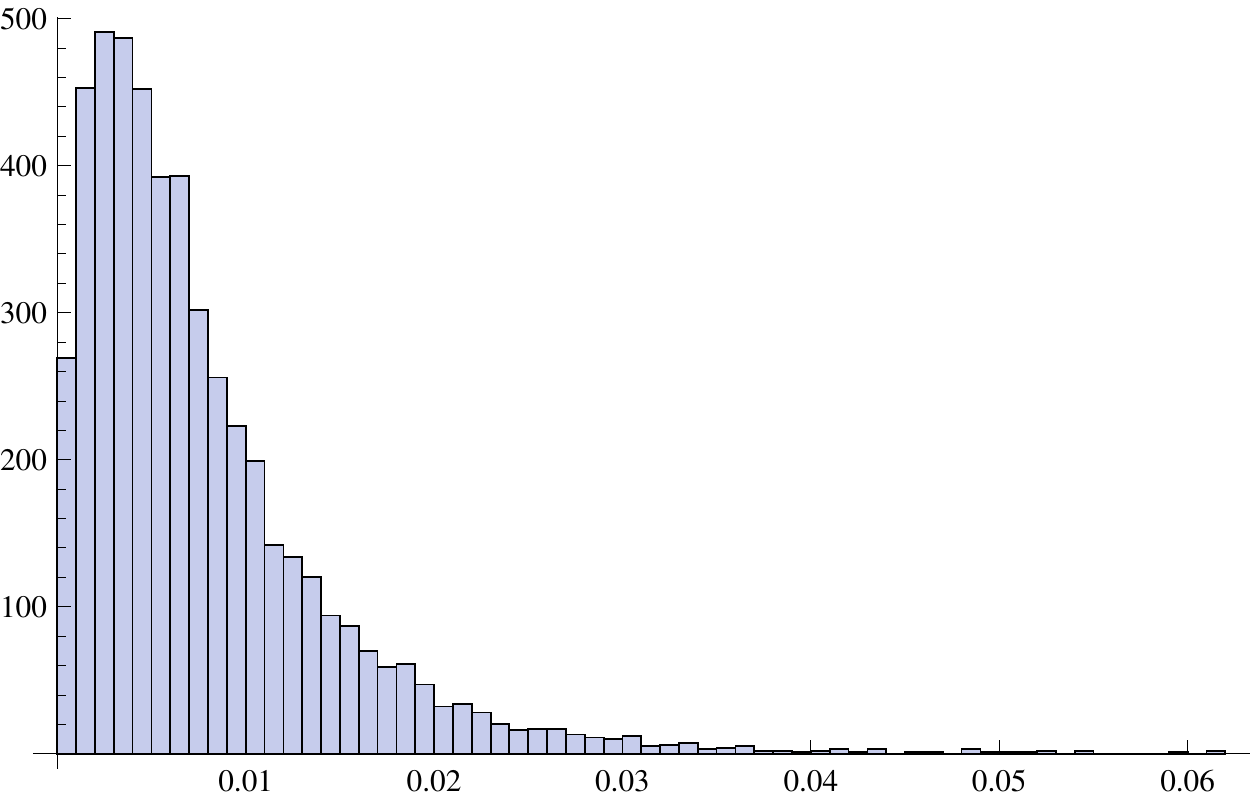}
  \caption{Left: a typical flux tube event from the decay of a 30 GeV flux tube in the toy model described in the text.
   Right: Sphericity distribution from 5000 decays of a 30 GeV flux tube in the toy model described in the
   text.}\label{fig:fluxtubeevent}
\end{figure}

For convenience we use the same string tension as in Ref.
\cite{Casher:1978wy}, $\sigma = 0.177~{\rm GeV}^2$. We simulated a
flux tube initially consisting of a quark and antiquark produced at
the origin moving with $p_z = \pm 15 {\rm GeV}$ along the $z$-axis.
The overall decay probability is suppressed from the original
calculations to take into account the effect of large $N$: for the
plots shown the probability per unit time and unit length of a break
is $\frac{p}{L \Delta T} = 0.01 {\rm GeV}^2$. We take time steps of
0.025 {\rm GeV}$^{-1}$. As a stopping criterion, after each decay we
flag as stable any new string segment that has \beq m =
\sqrt{E(|{\bf x}_i - {\bf x}_{i+1}|,p_i,p_{i+1})^2 - |{\bf p}_i +
{\bf p}_{i+1}|^2} < 1.54\,\,{\rm GeV}. \eeq As we can see in
Fig.~\ref{fig:fluxtubeevent} these events are quite jetty, with very
low characteristic sphericities, as expected in QCD.

\subsection{Including String Dynamics}\label{sec:model2}

We can move beyond this crude model of flux tubes by including
stringy dynamics: we will  model the flux tube worldsheet with the
classical Nambu-Goto action. We specialize to static gauge, $t =
\tau$, and parametrize the spatial coordinate $s$ with the conditions
\beq
\label{eq:stringpar1}
\partial_s \vec{X} \cdot \partial_\tau \vec{X} & = & 0, \\
 \label{eq:stringpar2}
\left(\partial_s \vec{X}\right)^2 + \left(\partial_\tau
\vec{X}\right)^2 & = & 1, \eeq which reduce the string action to
\beq\label{eq:SNG} S_{NG} = \frac{\sigma}{2}  \int d\tau ds \left(1
+ \left(\partial_s \vec{X}\right)^2 -  \left(\partial_\tau
\vec{X}\right)^2 \right), \eeq so that the equations of motion are
simple wave equations (for a detailed discussion, see the textbook
\cite{Zwiebach:2004tj}). Note that with this parametrization, the
string energy is simply the tension $\sigma$ times the length of the
$s$ interval.

There is a subtlety here: if we use just the Nambu-Goto action for
the string, with  Neumann boundary conditions, the string endpoints
will always move orthogonal to the string direction. However, we
begin with a string stretched along the $z$-axis between two quarks
with momentum in the $z$-direction. The Nambu-Goto action alone will
not suffice to describe this system. This puzzle is resolved by
noting that the quarks on the ends of the string are physical
particles, and we should add the action of a massive point particle
at each string endpoint to the
Nambu-Goto action, with the constraint that the string must end at
the location of the point particle.\footnote{Note that in the
AdS/CFT context, one deals with strings that are described by a
supersymmetric string action with $D$-brane boundary conditions
moving in some curved space. In particular, there is no added
point-particle action of the type we are considering here. However,
AdS/CFT can still describe open strings that are dual to a quark and
antiquark moving back-to-back with a flux tube forming in between.
The resolution of the puzzle in this context is that the AdS/CFT
dual involves a string that is falling in the AdS radial
(holographic) direction. For a more detailed discussion, including
some remarks about how AdS radial evolution can be dual to the
parton shower, see Ref. \cite{Strassler:2008bv}.} There is a recent
discussion of this in Ref. \cite{Quirks} in the context of
``quirks", hypothetical long strings of some new gauge group with
$M_Q \gg \Lambda$, which we will follow for the key points. As in
our simple Hamiltonian toy model, the quark (and antiquark) will
experience a force, but now we can have a proper relativistic and
local description: the force is related to the direction the quark
moves in, and the direction tangent to the string at its endpoint.
Specifically:
\beq
\label{eq:force} {\vec F} = -\sigma \left(\sqrt{1 -
{{\vec v}_\perp}^{\,2}} \frac{{\vec v}_\parallel}{v_\parallel} +
\frac{v_\parallel}{\sqrt{1 -{ {\vec v}_\perp}^{\,2}}} {\vec v}_\perp
\right),
\label{Nambu-Gotoeqom}
\eeq
where $v_\parallel$ is the component of quark velocity
along the string direction and ${\vec v}_\perp$ is the orthogonal
component. The corresponding rate of change of the quark energy
is
\beq \label{eq:quarkenergyloss}
\frac{dE_q}{dt} = -\sigma \frac{v_\parallel}{\sqrt{1-{\vec v}_\perp^{\,2}}}.
\eeq
 If this force is acting to slow down the quark, then
energy conservation will require that the quark leaves a new bit of
flux tube in its wake. If the quark is speeding up, then it is
eating some existing flux tube as it moves. Thus adding the point
particle action at the string boundary is, from a certain point of
view, an extreme deformation of the Nambu-Goto dynamics; we're not
just studying a classical string that is moving relativistically,
we're studying a string that grows and shrinks as it moves
relativistically. As noted above, the static gauge parametrization
that makes the string equation of motion simplest has the feature
that the length of the $s$ interval is the string energy in
string units. To continue to use that parametrization in this
context, we have to allow the size of the $s$ interval to grow
or shrink: if the string energy, in units of the tension, is
$\varepsilon(t) = E_{string}(t)/\sigma$, then we can use a
parametrization in which, for the initial back-to-back string,
 $s$ takes values in
$\left[-\varepsilon(t)/2, \varepsilon(t)/2\right]$. More generally,
after a string breaking, the quark and antiquark will lose energy at
different rates, so $s$ ranges in $\left[-\varepsilon_1(t),
\varepsilon_2(t)\right]$, with \beq
\frac{d\varepsilon_1}{dt} &=& -\frac{1}{\sigma} \frac{dE_q}{dt} \\
\frac{d\varepsilon_2}{dt} &=& -\frac{1}{\sigma} \frac{dE_{\bar q}}{dt}.
\eeq
We will
denote $\varepsilon_1(t) + \varepsilon_2(t)$ simply by  $\varepsilon(t)$,
which measures the string energy in units of $\sigma$.

Numerically, it's more convenient to use a coordinate living in a fixed interval,
$\tilde{s} = \frac{2 (s + \varepsilon_1(t))}{\varepsilon(t)} -1 \in [-1,1]$.
Because this new coordinate is a function of both the original
$s$ and $t = \tau$, the action will become more complicated:
what was the derivative with respect to $\tau$ at fixed $s$,
for instance, is now a combination of derivatives with respect to
$\tilde{s}$ and $\tau$:
\beq
\left.\frac{d{\vec X}}{d \tau}\right|_{s} & =   & \left.\frac{d{\vec X}}{d \tilde{s}}\right|_{\tau} \frac{d{\tilde s}}{dt} +  \left.\frac{d{\vec X}}{d \tau}\right|_{\tilde{s}}. \\
\frac{d{\tilde s}}{dt} & = & -2 \frac{d\varepsilon}{dt}
\frac{{\tilde s + 1}}{\varepsilon} + \frac{2}{\varepsilon}
\frac{d\varepsilon_1}{dt}. \eeq Note that $\frac{d{\tilde s}}{dt}$
is interpolating between the rate of quark and antiquark energy
change, in appropriate units, as we move along the string. Also note
that the term $\left.\frac{d{\vec X}}{d \tilde{s}}\right|_{\tau}
\frac{d{\tilde s}}{dt}$ accounts for the fact that the quark can
have a velocity component $v_\parallel$ along the string direction,
even though the string itself can carry only transverse momentum.

Thus we can rewrite the action (\ref{eq:SNG}) in the
$(\tilde{s},\tau)$  coordinate frame. It's somewhat ugly, as it
involves $\varepsilon_{1,2}(\tau)$ and their time-derivatives, so in this
parametrization there appears to be some nonlocal dependence on what
is happening at the string endpoints. Nonetheless, the key point is that
the change in string energy is entirely determined by the change in
energy of the quark and antiquark, as in Eq. (\ref{eq:quarkenergyloss}).
Thus, numerically, we have a simple procedure
to follow. We keep track of a fixed number $N_{bit}$ of ``string
bits," i.e. locations at fixed intervals in $\tilde{s}$ along
the string. In a small time interval $dt$, we can first calculate
from Eqns. (\ref{eq:force}) and (\ref{eq:quarkenergyloss})
 the change in momentum and energy of the
quark and antiquark at the ends of a string. This then tells us how
$\varepsilon(t)$ is changing. We then use the equations of motion
computed from the action Eq. (\ref{eq:SNG}), written in the
$(\tilde{s},\tau)$ frame and discretized, to evolve each
``string bit" forward $dt$ in time. In practice, we do this time
evolution with a fourth-order Runge-Kutta algorithm. We then make
small adjustments to the positions ${\vec X}(\tilde{s})$ and
velocities $\partial_\tau {\vec X}(\tilde{s})$ of the string
bits so that the parametrization conditions, Eqns.
(\ref{eq:stringpar1}) and (\ref{eq:stringpar2}), are satisfied as
precisely as possible numerically, because in practice small
discretization errors in these conditions seem more numerically
troublesome than discretization errors in the equations of motion
themselves. This gives us a slightly crude, but fully relativistic,
algorithm for evolving a configuration of strings forward in time.

The next step is to decay the strings. Here we use an {\em ad hoc}
modification of  the procedure from Section \ref{sec:model1}. We
continue to be guided by the Schwinger calculation, but we are now
taking it somewhat further from its domain of validity. Namely, we
assume that at a given instant in time, the decay probability of the
string is proportional to the length of the string,
and that the decay is equally likely to happen
anywhere in the coordinate $\tilde{s}$. Rather than transverse
momenta, we pull an energy from a Gaussian distribution,
$dN(E) \sim d^2
E~\exp\left(-\frac{\pi E^2}{\sigma}\right)$, insisting that $E > 2 m_q$,
and remove a length of $\tilde{s}$ interval that balances this energy.
In general this interval may also have some momentum. To construct
the momenta of the new quark and antiquark, we boost to the rest frame
of the removed string interval and construct the momenta of two daughters
of mass $m_q$, perpendicular to the string direction at the center of
the removed interval. (This ensures that in the case of the first breaking,
this algorithm matches the simple semiclassical flux tube algorithm of
our previous discussion.) Boosting back, we compute the components
$v_\parallel$ and $v_\perp$ of the velocities of the new $q$ and ${\bar q}$
and the corresponding rate of energy loss. We divide each
daughter string into $N_{bit}$ bits, just as the parent string had,
interpolating between bits of the parent string to obtain the
positions and velocities of the new bits. We place the final quark
and antiquark at a location such that the constraint (\ref{eq:stringpar2})
is satisfied.

\begin{figure}[!h]
  \centering
  \includegraphics[width = 0.5\textwidth]{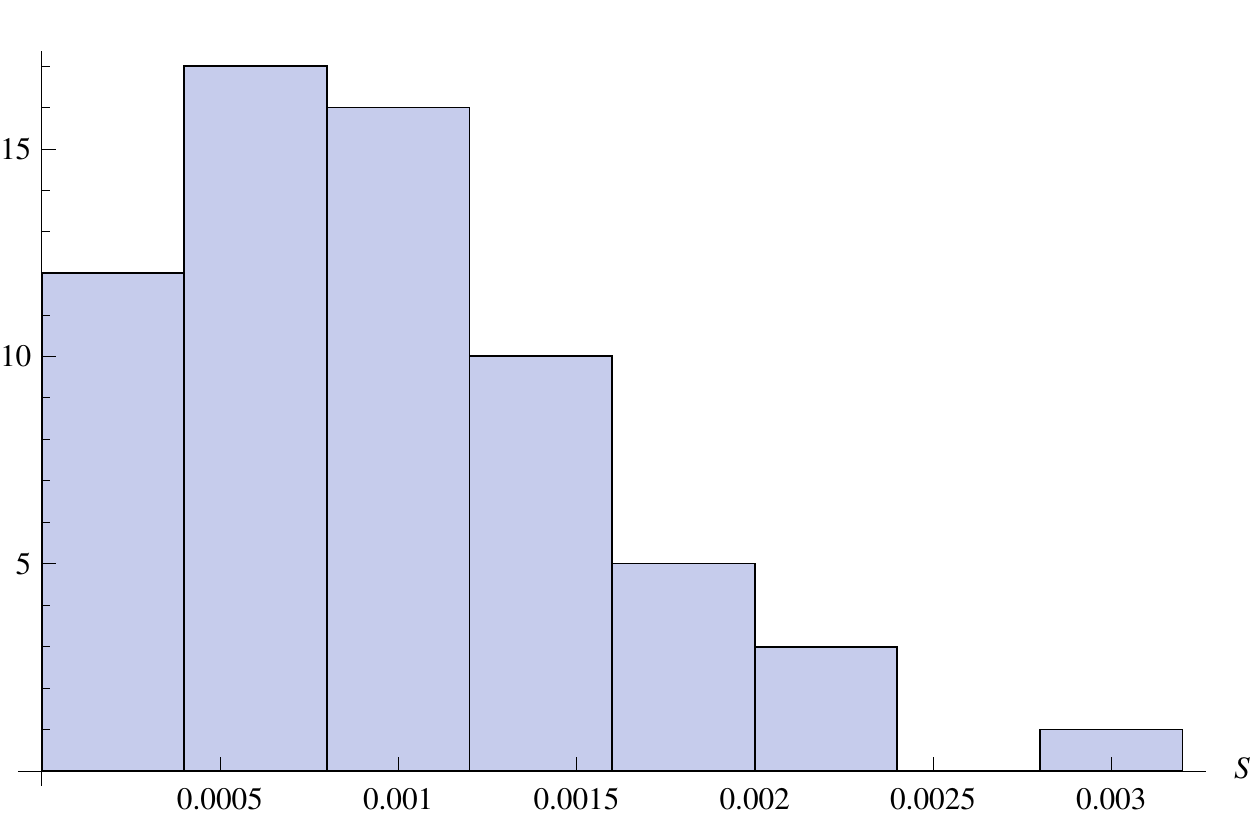}
  \caption{Sphericity distribution from the string model, after the first four string decays}\label{fig:stringspher}
\end{figure}

The numerics of this procedure prove to be tricky, and we present
results only for the first four decay steps.
The results of this model are similar to those of the simpler static
flux tube model.  We have fixed $N_{bit} = 101$ for the number of
points we track along the string, and have taken time steps of size
$0.003125$ GeV$^{-1}$. The quark mass is taken to be 0.3 GeV.
Other parameters (tension, initial $p_z$,
decay probability per unit time and unit length in string units) are
as in the previous subsection. In Figure \ref{fig:stringspher} we plot
the sphericity distribution. It is peaked toward low values, showing
that the string simulation predicts highly directional, jetty events
just as our simpler flux tube model did.

%%%%%%%%%%%%%%%%%%%%%%%%%%%%%%%%%%%%%%%%%%%%%%%%%%%%%%
\section{Conclusions}
\label{sec:conclude}
\setcounter{equation}{0}
\setcounter{footnote}{0}

We have examined various aspects of large $N$ gauge theories for
small and large 't Hooft coupling $\lambda$. We found that for small
't Hooft coupling agreement with perturbation theory suggests that
the mass spectrum should asymptotically show the characteristic
Regge-type behavior rather than the KK-type scaling, that arises in the
 large $\lambda$ limit. We have given a
simple toy model that sheds light on how one can interpolate between
these two spectra. This model suggests that the KK-like modes are
more likely to be deeply bound mesons in the Coulombic
short-distance part of the potential rather than genuine QCD-like
modes related to the confining long distance potential. The event
shapes resulting from scatterings in the small and large $\lambda$
theories also seem to be qualitatively different. While for small
$\lambda$ one expects QCD-like events forming jets, for large
$\lambda$ we have shown that the final states are much more
spherical, in agreement with the conjecture of Strassler
and the calculations of Hofman and Maldacena
for events in gravity duals of CFT's. We conclude that in order to
fully capture the dynamics of QCD with small $\lambda$ inclusion of
some sort of stringy dynamics seems to be unavoidable.

%%%%%%%%%%%%%%%%%%%%%%%%%%%%%%%%%%%%%%%%%%%%%%%%%%%%%%
%%%%%%%%%%%%%%%%%%%%%%%%%%%%%%%%%%%%%%%%%%%%%%%%%%%%%%

%%%%%%%%%%%%%%%%%%%%%%%%%%
\section*{Acknowledgments}
We thank I. Klebanov, M. Luty, E. Pajer, and J. Thaler for useful
discussions, and I. Klebanov and L. McAllister for comments on the manuscript.
We thank the Aspen Center for Physics and the KITP in
Santa Barbara for hospitality during the completion of this work.
The research of C.C. is partly supported by the NSF under grant
PHY-0355005 and by a US-Israeli BSF grant. J.T. is
supported in part by the US Department of Energy under contract No.
DE- FG03-91ER40674 and by a US-Israeli BSF grant.

\appendix

\section*{Appendix}

\section{Towards the completion of Migdal's program}

In this appendix we describe an attempt at completing Migdal's
program of analytically continuing the PT results for QCD into
meromorphic functions valid for all energy regimes, along the lines
of discussion in Sec.~\ref{sec:mero}. This will lead us toward functions
whose pole structure seems to asymptotically agree with that of the
digamma function, however as we will see some other more exotic
possibilities can not be completely ruled out based on our
arguments.

Let us try to find a function that is a good fit to PT both in the
deep Euclidean regime and the Minkowski regime away from the real
axis (eg. also matches $\rho_{\Delta}$ for $\Delta \sim 1$). We will
be looking for a meromorphic function of the form
\begin{equation}
\Pi_{MR} (s)=\sum_{n=1}^\infty \frac{r_n}{s-m_n^2}.
\end{equation}
We require that this function (just as the Pad\'e approximant)
reproduces $\log s$ for $-s \gg 1$ with no $1/s$ corrections. But in
addition we also require that it be a good fit to $\rho_{\Delta}(s)$
for $s>1$. Let us first understand why the Pad\'e approximation
fails this requirement. The problem with the Pad\'e approximation is
that the succession of poles and residues obey $m_n^2 \sim n^2$ and
also $r_n \sim n$. The expression for $\rho_{\Delta}$ is
\begin{equation}
\rho_{\Delta} (s) = \sum_n \frac{r_n
\Delta^2}{(s-m_n^2)^2+\Delta^2}.
\end{equation}
Clearly, for $s=m_n^2$ one gets a contribution of $r_n$, while the
other terms are highly suppressed by the denominator. If $|r_n|$ is
not asymptotically approaching a constant (or decreasing), then
there is no way to correctly approximate for large $n$ the
asymptotically constant jump across the branch cut of the logarithm.
Thus we conclude that the distribution of residues should be such
that for large $n$ $r_n$ should be asymptotically bounded from
above. This is the condition that the Pad\'e approximant is not
satisfying.

Now, we claim that this condition on the $r_n$ in fact prohibits any
function where the separation between poles grows asymptotically,
not just the Pad\'e approximant. For example, there is no function
with a similar mass pattern $m_n^2 \sim n^2$ but with nonincreasing
residues that provides a match to PT. This is
intuitively clear: adding together functions supported on a narrow
interval, growing progressively further apart, will never produce a
constant. We can formalize this: take the average value of
$\rho_\Delta$ between two consecutive poles $m_n^2$ and $m_{n+1}^2$.
If $\rho_\Delta$ is to agree with PT, this average
should not go to zero as $n \rightarrow \infty$. But we can just
compute the average:
\begin{equation}
\bar{\rho}_{\Delta,n} \equiv \frac{1}{m_{n+1}^2 - m_n^2}
\int^{m_{n+1}^2}_{m_n^2} ds~ \sum_k \frac{r_k \Delta^2}{(s -
m_k^2)^2 + \Delta^2}.
\end{equation}
Interchanging the order of summation and integration, this is just
\begin{equation} \label{eq:rhoave}
\bar{\rho}_{\Delta,n} = \frac{1}{m_{n+1}^2 - m_n^2} \sum_k
\frac{r_k}{\Delta} \left(\arctan\frac{m_k^2 - m_{n+1}^2}{\Delta} -
\arctan\frac{m_k^2-m_n^2}{\Delta}\right).
\end{equation}
This expression is valid in general, regardless of the behavior of
the $m_n^2$ and $r_n$.

Now we exploit the properties of the arctangent. If the distance
between successive poles is asymptotically growing, we can choose
$n$ such that $\frac{m_{n+1}^2 - m_n^2}{\Delta} \gg 1$. Then we use
the fact that:
\begin{equation}
\arctan(x) - \arctan(x - x_0) \approx \pi \theta(x) \theta(x_0 -
x),~~x_0 \gg 1.
\end{equation}
In other words, for very large values of $x_0$ this function is
approximately $\pi$ on the interval  from $0$ to $x_0$ and 0 outside
that interval. In particular, in our case we identify $x$ with
$\frac{m_k^2 - m_{n+1}^2}{\Delta}$ and $x_0$ with $\frac{m_{n+1}^2 -
m_n^2}{\Delta}$. It is then clear that the only $k$ values that
contribute to the sum in Eq.~\ref{eq:rhoave} are $k = n+1$ and $k =
n$. The result is that
\begin{equation}
\bar{\rho}_{\Delta,n} \approx  \frac{1}{m_{n+1}^2 - m_n^2}
\frac{(r_n + r_{n+1})\arctan(m_{n+1}^2 - m_n^2)}{\Delta}.
\end{equation}
We have argued that the $r_n$ are not asymptotically growing; the
arctangent is also bounded (it is  approximately $\pi$). Hence, if
the denominator $m_{n+1}^2 - m_n^2$ is growing, this expression
tends asymptotically to zero at large $n$, in contradiction with
PT.

At this point we have established that the radial Regge physics,
$m_n^2 \sim n$ asymptotically,  is the behavior which matches
PT with the fewest poles on a given interval. We do
not have an argument that we cannot have slower growth, for instance
$m_n^2 \sim \sqrt{n}$. Intuitively it is apparent that if we {\em
do} have slower growth, the residues must decrease or we will
construct a function that greatly exceeds the perturbative answer at
large $s$. Let's see what happens if we attempt to extend the above
argument to the case of poles that grow successively closer. In that
case we are interested in the opposite limit of the difference of
arctangents, $x_0 \rightarrow 0$. This is of course determined by
the derivative:
\begin{equation}
\arctan(x) - \arctan(x-x_0) \approx \frac{x_0}{x^2 + 1},~~x_0 \ll 1.
\end{equation}
Assuming the separation between poles is decreasing with $n$, we can
choose $n$ such that $\frac{m_{n+1}^2 - m_n^2}{\Delta} \ll 1$ and
the above formula is valid. Note that $\frac{x_0}{x^2 + 1} >
\frac{x_0}{2}$ provided $|x| < 1$. Again identifying $x$ with
$\frac{m_k^2 - m_{n+1}^2}{\Delta}$ and summing only over the terms
where $|x| < 1$, we find that the sum is {\em at least}
\begin{equation}
\bar{\rho}_{\Delta,n} \gtsimeq \frac{1}{m_{n+1}^2-m_n^2}
\sum_{k:~|m_n^2 - m_k^2| < \Delta} \frac{r_k}{\Delta}
\frac{m_{n+1}^2-m_n^2}{2 \Delta} = \frac{1}{2 \Delta^2}
\sum_{k:~|m_n^2 - m_k^2| < \Delta}  r_k.
\end{equation}
From this we get an estimate of how quickly the residues $r_k$ must
decrease: fast enough to  compensate for the number of poles in a
given interval. Assuming $m_n^2$ is a smooth function of $n$, for
large $n$ the number of poles in a unit interval is characterized by
$(m_{n+1}^2 - m_n^2)^{-1}$, so the residue should satisfy $r_n
\approx m_{n+1}^2 - m_n^2$.

Now, since we don't have any argument against decreasing poles, it
seems possible that a function like
\begin{equation}
\Pi(s) = \sum_n \left(\frac{1}{\sqrt{n}(s-\sqrt{n})} +
\frac{1}{n}\right)
\end{equation}
could provide a good match to PT. We have
numerically investigated this, and indeed such a function appears
to agree well. It does
not seem possible to determine the correct behavior of the
distribution of poles just from matching the leading perturbative
behavior of two-point functions. On the other hand, the
asymptotically constant distribution of pole expected from Regge
physics is in some sense the {\em minimal} choice, so in the absence
of other arguments it seems preferable.

\section{Lorentz Invariance of Flux tube Breaking}

Here we show that for center of mass energies much larger than the the typical
transverse momentum $\sqrt{\sigma}$, the simple model of Section \ref{sec:model1}
gives a Lorentz invariant description of the early decays of a flux tube.  At large enough
center of mass energies the flux tube stretches in a straight line between the back-to-back  quark and anti-quark, and the transverse momenta are negligible.
In this limit, the situation reduces to a 1+1 dimensional problem.  The discussion can be further simplified
by noting that the pair production  which breaks the flux tube is just like 2D bubble nucleation, which is Lorentz invariant around the center of the bubble \cite{bubbles,fluxbubble,barger}.

Let us take the string axis to be along the
$x$ direction, and our string breaking pair production to be centered at $x=0$, $t=0$, which means that the original production of the flux tube took place at some earlier (negative) time.  At $t=0$, we will take a quark of mass $m$ to be at $x=x_0$ and moving in the positive $x$ direction with momentum $p_0$; the anti-quark at the other end of the flux tube is somewhere along the negative $x$ axis moving in the negative $x$ direction.  The Nambu-Goto-quark equations of motion (\ref{Nambu-Gotoeqom}) tell us that the momentum of the quark will decrease linearly with time:
\beq
p_q=p_0-\sigma t~.
\eeq
When the flux tube breaks, a new quark---anti-quark pair is formed at $t=0$.  Neglecting the small transverse momentum, the new pair is at rest, and in order to conserve energy, a section of the flux tube must be removed to account for their rest energy.  Thus the new anti-quark of mass $m$ is at
\beq
x=d=m/\sigma
\eeq
while the new quark is at $x=-d$, so that the flux tube section from $-d$ to $d$ has donated its energy
to provide the new masses. Again the Nambu-Goto-quark equations of motion (\ref{Nambu-Gotoeqom}) tell us
that the subsequent momentum of the new anti-quark is given by
\beq
p_a=\sigma t~.
\eeq
Integrating the velocity, $v_a=p_a/E_a$,  of the new anti-quark we find that its subsequent position is
\beq
x_a=\int v_a \,dt = \int \frac{p_a}{\sqrt{m^2+p_a^2} }\,dt=\sqrt{d^2+t^2}~.
\eeq
This is, of course, exactly what we expect from 2D bubble nucleation: the trajectory should be an hyperboloid, corresponding to a circle in Euclidean space ($x^2+t_E^2=d^2$).  Similarily, the trajectory of the original quark can be found by integration as well but the result just corresponds to another  circle  in Euclidean space, $(x-x_c)^2+(t_E-t_c)^2=d^2$, however one that is not centered about the origin.  The center is given by
\beq
x_c &=& x_0+\frac{\sqrt{m^2+p_0^2}}{\sigma} =x_0+\sqrt{d^2+\frac{p_0^2}{\sigma^2} }~,\\
t_c &=& \frac{p_0}{\sigma}~.
\eeq
Thus the position of the quark is simply given by
\beq
x_q=x_c-\sqrt{d^2+(t_c-t)^2}~.
\eeq

Now let us calculate the rest mass of the daughter flux tube.  This is simple at $t=0$.  The total momentum is just that of the quark (since a string cannot carry longitudinal momentum), so $p=p_0$.  The total energy is
\beq
E= m +\sigma(x_0-d) +\sqrt{m^2+p_0^2}=\sigma\, x_0 +\sqrt{m^2+p_0^2}~.
\label{totalenergy}
\eeq
Thus the rest mass is
\beq
m_{rest}=\sqrt{E^2-p^2}=\sqrt{m^2+\sigma^2 x_0^2+2\sigma x_0 \sqrt{m^2+p_0^2}}~.
\eeq

What if we had chosen to calculate the rest mass in some other frame? The issue may seem quite complex since a boost will change the length of the string, but  the string length is changing with time so  the new time slicing will actually involve the string ends at different stages of evolution.  However the situation is really quite simple.  If we boost to a frame moving with velocity $v$ to the right, special relativity tells us that in that frame the energy and momentum will be
\beq
E^\prime=\gamma(E- v p_0); \quad p^\prime=\gamma(p_0-v E)~.
\label{sr}
\eeq
Since the new anti-quark trajectory is boost invariant it will be at $x^\prime=d$ at $t^\prime=0$, so in analogy to Eq. (\ref{totalenergy}) we should also have that at $t^\prime=0$ the total energy in this frame is
\beq
E^\prime=\sigma \, x_q^\prime +\sqrt{m^2+p^{\prime 2}}~,
\label{totalenergyprime}
\eeq
where $x_q^\prime$ is the position of the quark in the boosted frame at $t^\prime=0$. Comparing Eqs. (\ref{sr}) and (\ref{totalenergyprime}) tells us what $x_q^\prime$ should be if the model is Lorentz invariant.  We can independently calculate $x_q^\prime$ from the boost of the quark trajectory.  This is most easily done by rotating the center of the corresponding circle in Euclidean space (which is equivalent to boosting the the focus of the hyperboloid in Minkowski space).  This gives, at $t^\prime=0$,
\beq
x_q^\prime=\gamma\left(x_c- v \,t_c \right)-\sqrt{d^2+\gamma^2\left(
t_c -v \,x_c \right)^2}~,
\eeq
which is exactly what is required for Eqs. (\ref{sr}) and (\ref{totalenergyprime}) to agree.
Thus the rest mass of the daughter flux tube is the same in any boosted frame, and the model is indeed Lorentz invariant.


\begin{thebibliography}{99}

\bibitem{maldacena}
J.~M.~Maldacena,
  %``The large N limit of superconformal field theories and supergravity,''
  Adv.\ Theor.\ Math.\ Phys.\  {\bf 2}, 231 (1998)
  [Int.\ J.\ Theor.\ Phys.\  {\bf 38}, 1113 (1999)]
  [ {\tt hep-th/9711200}];
  %%CITATION = IJTPB,38,1113;%%
    S.~S.~Gubser, I.~R.~Klebanov and A.~M.~Polyakov,
  %``Gauge theory correlators from non-critical string theory,''
  Phys.\ Lett.\  B {\bf 428}, 105 (1998)
  [arXiv:hep-th/9802109].
  %%CITATION = PHLTA,B428,105;%%
E.~Witten,
  %``Anti-de Sitter space and holography,''
  Adv.\ Theor.\ Math.\ Phys.\  {\bf 2}, 253 (1998)
  [ {\tt hep-th/9802150}].
  %%CITATION = 00203,2,253;%%

\bibitem{AdSQCD}
 J.~Erlich, E.~Katz, D.~T.~Son and M.~A.~Stephanov,
  %``QCD and a holographic model of hadrons,''
  {\tt hep-ph/0501128};
  %%CITATION = HEP-PH 0501128;%%
 L.~Da Rold and A.~Pomarol,
  %``Chiral symmetry breaking from five dimensional spaces,''
  {\tt hep-ph/0501218}.
  %%CITATION = HEP-PH 0501218;%%

\bibitem{Strassler:2008bv}
  M.~J.~Strassler,
  %``Why Unparticle Models with Mass Gaps are Examples of Hidden Valleys,''
  arXiv:0801.0629 [hep-ph].
  %%CITATION = ARXIV:0801.0629;%%

\bibitem{PolchinskiStrasslerDIS}
  J.~Polchinski and M.~J.~Strassler,
  %``Deep inelastic scattering and gauge/string duality,''
  JHEP {\bf 0305}, 012 (2003)
  [arXiv:hep-th/0209211].
  %%CITATION = JHEPA,0305,012;%%

\bibitem{HofmanMaldacena}
  D.~M.~Hofman and J.~Maldacena,
  %``Conformal collider physics: Energy and charge correlations,''
  JHEP {\bf 0805}, 012 (2008)
  [arXiv:0803.1467 [hep-th]].
  %%CITATION = JHEPA,0805,012;%%

\bibitem{HIM}
  Y.~Hatta, E.~Iancu and A.~H.~Mueller,
  %``Jet evolution in the N=4 SYM plasma at strong coupling,''
  JHEP {\bf 0805}, 037 (2008)
  [arXiv:0803.2481 [hep-th]]
  %%CITATION = JHEPA,0805,037;%%

\bibitem{MoreHatta}
  Y.~Hatta, E.~Iancu and A.~H.~Mueller,
  %``Deep inelastic scattering off a N=4 SYM plasma at strong coupling,''
  JHEP {\bf 0801}, 063 (2008)
  [arXiv:0710.5297 [hep-th]];
  %%CITATION = JHEPA,0801,063;%%
   Y.~Hatta and T.~Matsuo,
  %``Jet fragmentation and gauge/string duality,''
  arXiv:0804.4733 [hep-th];
  %%CITATION = ARXIV:0804.4733;%%
  Y.~Hatta and T.~Matsuo,
  %``Thermal hadron spectrum in $e^+e^-$ annihilation from gauge/string
  %duality,''
  arXiv:0807.0098 [hep-ph];
  %%CITATION = ARXIV:0807.0098;%%
  Y.~Hatta,
  %``Relating $e^+e^-$ annihilation to high energy scattering at weak and strong
  %coupling,''
  arXiv:0810.0889 [hep-ph].
  %%CITATION = ARXIV:0810.0889;%%

\bibitem{'tHooft:1973jz}
  G.~'t Hooft,
  %``A PLANAR DIAGRAM THEORY FOR STRONG INTERACTIONS,''
  Nucl.\ Phys.\  B {\bf 72}, 461 (1974).
  %%CITATION = NUPHA,B72,461;%%

\bibitem{RS}
  L.~Randall and R.~Sundrum,
  %``A large mass hierarchy from a small extra dimension,''
  Phys.\ Rev.\ Lett.\  {\bf 83}, 3370 (1999)
  [arXiv:hep-ph/9905221].
  %%CITATION = PRLTA,83,3370;%%

\bibitem{KS}
  I.~R.~Klebanov and M.~J.~Strassler,
  %``Supergravity and a confining gauge theory: Duality cascades and
  %chiSB-resolution of naked singularities,''
  JHEP {\bf 0008}, 052 (2000)
  [arXiv:hep-th/0007191].
  %%CITATION = JHEPA,0008,052;%%

  \bibitem{'tHooft:1977hy}
  G.~'t Hooft,
  %``On The Phase Transition Towards Permanent Quark Confinement,''
  Nucl.\ Phys.\  B {\bf 138}, 1 (1978).
  %%CITATION = NUPHA,B138,1;%%

 \bibitem{confinement}
   K.~G.~Wilson,
  %``CONFINEMENT OF QUARKS,''
  Phys.\ Rev.\  D {\bf 10}, 2445 (1974);
  %%CITATION = PHRVA,D10,2445;%%
    Y.~Nambu,
  %``Strings, monopoles, and gauge fields,''
  Phys.\ Rev.\  D {\bf 10}, 4262 (1974).
  %%CITATION = PHRVA,D10,4262;%%
   Y.~Nambu,
  %``QCD And The String Model,''
  Phys.\ Lett.\  B {\bf 80}, 372 (1979);
  %%CITATION = PHLTA,B80,372;%%
    M.~Luscher, G.~Munster and P.~Weisz,
  %``How Thick Are Chromoelectric Flux Tubes?,''
  Nucl.\ Phys.\  B {\bf 180}, 1 (1981);
  %%CITATION = NUPHA,B180,1;%%
   M.~Luscher,
  %``Symmetry Breaking Aspects Of The Roughening Transition In Gauge Theories,''
  Nucl.\ Phys.\  B {\bf 180}, 317 (1981);
  %%CITATION = NUPHA,B180,317;%%
  R.~Sundrum,
  %``Hadronic string from confinement,''
  arXiv:hep-ph/9702306.
  %%CITATION = HEP-PH/9702306;%%

 \bibitem{stringyRegge}
  A.~Armoni, J.~L.~F.~Barbon and A.~C.~Petkou,
  %``Rotating strings in confining AdS/CFT backgrounds,''
  JHEP {\bf 0210}, 069 (2002)
  [arXiv:hep-th/0209224];
  %%CITATION = JHEPA,0210,069;%%
  L.~A.~Pando Zayas, J.~Sonnenschein and D.~Vaman,
  %``Regge trajectories revisited in the gauge / string correspondence,''
  Nucl.\ Phys.\  B {\bf 682}, 3 (2004)
  [arXiv:hep-th/0311190];
  %%CITATION = NUPHA,B682,3;%%
    F.~Bigazzi, A.~L.~Cotrone, L.~Martucci and L.~A.~Pando Zayas,
  %``Wilson loop, Regge trajectory and hadron masses in a Yang-Mills theory
  %from semiclassical strings,''
  Phys.\ Rev.\  D {\bf 71}, 066002 (2005)
  [arXiv:hep-th/0409205];
  %%CITATION = PHRVA,D71,066002;%%
   M.~Kruczenski, L.~A.~P.~Zayas, J.~Sonnenschein and D.~Vaman,
  %``Regge trajectories for mesons in the holographic dual of large-N(c)  QCD,''
  JHEP {\bf 0506}, 046 (2005)
  [arXiv:hep-th/0410035].
  %%CITATION = JHEPA,0506,046;%%

\bibitem{PRT}
  J.~M.~Pons, J.~G.~Russo and P.~Talavera,
  %``Semiclassical string spectrum in a string model dual to large N QCD,''
  Nucl.\ Phys.\  B {\bf 700}, 71 (2004)
  [arXiv:hep-th/0406266].
  %%CITATION = NUPHA,B700,71;%%

\bibitem{WittenBH}
  E.~Witten,
  %``Anti-de Sitter space, thermal phase transition, and confinement in  gauge
  %theories,''
  Adv.\ Theor.\ Math.\ Phys.\  {\bf 2}, 505 (1998)
  [arXiv:hep-th/9803131].
  %%CITATION = 00203,2,505;%%

\bibitem{KKmodes}
  C.~Csaki, H.~Ooguri, Y.~Oz and J.~Terning,
  %``Glueball mass spectrum from supergravity,''
  JHEP {\bf 9901}, 017 (1999)
  [arXiv:hep-th/9806021];
  %%CITATION = JHEPA,9901,017;%%
  H.~Ooguri, H.~Robins and J.~Tannenhauser,
  %``Glueballs and their Kaluza-Klein cousins,''
  Phys.\ Lett.\  B {\bf 437}, 77 (1998)
  [arXiv:hep-th/9806171].
  %%CITATION = PHLTA,B437,77;%%

\bibitem{MesonSpect}
    M.~Kruczenski, D.~Mateos, R.~C.~Myers and D.~J.~Winters,
  %``Meson spectroscopy in AdS/CFT with flavour,''
  JHEP {\bf 0307}, 049 (2003)
  [arXiv:hep-th/0304032].
  %%CITATION = JHEPA,0307,049;%%
  
 \bibitem{quarkonium5D}
    S.~Hong, S.~Yoon and M.~J.~Strassler,
  %``Quarkonium from the fifth dimension,''
  JHEP {\bf 0404}, 046 (2004)
  [arXiv:hep-th/0312071].
  %%CITATION = JHEPA,0404,046;%%

\bibitem{AdSMesons}
  T.~Sakai and J.~Sonnenschein,
  %``Probing flavored mesons of confining gauge theories by supergravity,''
  JHEP {\bf 0309}, 047 (2003)
  [arXiv:hep-th/0305049];
  %%CITATION = JHEPA,0309,047;%%
    J.~Babington, J.~Erdmenger, N.~J.~Evans, Z.~Guralnik and I.~Kirsch,
  %``Chiral symmetry breaking and pions in non-supersymmetric gauge /  gravity
  %duals,''
  Phys.\ Rev.\  D {\bf 69}, 066007 (2004)
  [arXiv:hep-th/0306018];
  %%CITATION = PHRVA,D69,066007;%%
  S.~Hong, S.~Yoon and M.~J.~Strassler,
  %``On the couplings of vector mesons in AdS/QCD,''
  JHEP {\bf 0604}, 003 (2006)
  [arXiv:hep-th/0409118];
  %%CITATION = JHEPA,0604,003;%%
  J.~Erdmenger, N.~Evans, I.~Kirsch and E.~Threlfall,
  %``Mesons in Gauge/Gravity Duals - A Review,''
  Eur.\ Phys.\ J.\  A {\bf 35}, 81 (2008)
  [arXiv:0711.4467 [hep-th]].
  %%CITATION = EPHJA,A35,81;%%

\bibitem{KSKK}
  M.~Berg, M.~Haack and W.~Mueck,
  %``Glueballs vs. gluinoballs: Fluctuation spectra in non-AdS/non-CFT,''
  Nucl.\ Phys.\  B {\bf 789}, 1 (2008)
  [arXiv:hep-th/0612224];
  %%CITATION = NUPHA,B789,1;%%
    M.~K.~Benna, A.~Dymarsky, I.~R.~Klebanov and A.~Solovyov,
  %``On Normal Modes of a Warped Throat,''
  JHEP {\bf 0806}, 070 (2008)
  [arXiv:0712.4404 [hep-th]].
  %%CITATION = JHEPA,0806,070;%%
  
\bibitem{deconstructed}
  D.~T.~Son and M.~A.~Stephanov,
  %``QCD and dimensional deconstruction,''
  Phys.\ Rev.\  D {\bf 69}, 065020 (2004)
  [arXiv:hep-ph/0304182].
  %%CITATION = PHRVA,D69,065020;%%

 \bibitem{annulons}
    E.~G.~Gimon, L.~A.~Pando Zayas, J.~Sonnenschein and M.~J.~Strassler,
  %``A soluble string theory of hadrons,''
  JHEP {\bf 0305}, 039 (2003)
  [arXiv:hep-th/0212061].
  %%CITATION = JHEPA,0305,039;%%

 \bibitem{adjointtrap}
   S.~Hong, S.~Yoon and M.~J.~Strassler,
  %``Adjoint trapping: A new phenomenon at strong 't Hooft coupling,''
  JHEP {\bf 0603}, 012 (2006)
  [arXiv:hep-th/0410080].
  %%CITATION = JHEPA,0603,012;%%
  
 \bibitem{PlasmaBalls}
   O.~Aharony, S.~Minwalla and T.~Wiseman,
  %``Plasma-balls in large N gauge theories and localized black holes,''
  Class.\ Quant.\ Grav.\  {\bf 23}, 2171 (2006)
  [arXiv:hep-th/0507219].
  %%CITATION = CQGRD,23,2171;%%

\bibitem{N4FluxTubes}
  I.~R.~Klebanov, J.~M.~Maldacena and C.~B.~.~Thorn,
  %``Dynamics of flux tubes in large N gauge theories,''
  JHEP {\bf 0604}, 024 (2006)
  [arXiv:hep-th/0602255];
  %%CITATION = JHEPA,0604,024;%%
  R.~C.~Brower, C.~I.~Tan and C.~B.~Thorn,
  %``String / flux tube duality on the lightcone,''
  Phys.\ Rev.\  D {\bf 73}, 124037 (2006)
  [arXiv:hep-th/0603256].
  %%CITATION = PHRVA,D73,124037;%%

 \bibitem{PolchinskiStrassler}
   J.~Polchinski and M.~J.~Strassler,
  %``Hard scattering and gauge/string duality,''
  Phys.\ Rev.\ Lett.\  {\bf 88}, 031601 (2002)
  [arXiv:hep-th/0109174];
  %%CITATION = PRLTA,88,031601;%%
 R.~C.~Brower, J.~Polchinski, M.~J.~Strassler and C.~I.~Tan,
  %``The Pomeron and Gauge/String Duality,''
  JHEP {\bf 0712}, 005 (2007)
  [arXiv:hep-th/0603115];
  %%CITATION = JHEPA,0712,005;%%
  R.~C.~Brower, M.~J.~Strassler and C.~I.~Tan,
  %``On The Pomeron at Large 't Hooft Coupling,''
  arXiv:0710.4378 [hep-th].
  %%CITATION = ARXIV:0710.4378;%%

 \bibitem{Eikonal}
   L.~Cornalba, M.~S.~Costa, J.~Penedones and R.~Schiappa,
  %``Eikonal approximation in AdS/CFT: From shock waves to four-point
  %functions,''
  JHEP {\bf 0708}, 019 (2007)
  [arXiv:hep-th/0611122];
  %%CITATION = JHEPA,0708,019;%%
  L.~Cornalba, M.~S.~Costa, J.~Penedones and R.~Schiappa,
  %``Eikonal approximation in AdS/CFT: Conformal partial waves and finite N
  %four-point functions,''
  Nucl.\ Phys.\  B {\bf 767}, 327 (2007)
  [arXiv:hep-th/0611123];
  %%CITATION = NUPHA,B767,327;%%
  L.~Cornalba, M.~S.~Costa and J.~Penedones,
  %``Eikonal Approximation in AdS/CFT: Resumming the Gravitational Loop
  %Expansion,''
  JHEP {\bf 0709}, 037 (2007)
  [arXiv:0707.0120 [hep-th]];
  %%CITATION = JHEPA,0709,037;%%
  L.~Cornalba, M.~S.~Costa and J.~Penedones,
  %``Eikonal Methods in AdS/CFT: BFKL Pomeron at Weak Coupling,''
  JHEP {\bf 0806}, 048 (2008)
  [arXiv:0801.3002 [hep-th]];
  %%CITATION = JHEPA,0806,048;%%
  L.~Cornalba and M.~S.~Costa,
  %``Saturation in Deep Inelastic Scattering from AdS/CFT,''
  arXiv:0804.1562 [hep-ph].
  %%CITATION = ARXIV:0804.1562;%%

\bibitem{Unsal}
  M.~Unsal,
  %``Quantum phase transitions and new scales in QCD-like theories,''
  arXiv:0807.0466 [hep-th].
  %%CITATION = ARXIV:0807.0466;%%

\bibitem{SakaiSugimoto}
  T.~Sakai and S.~Sugimoto,
  %``Low energy hadron physics in holographic QCD,''
  Prog.\ Theor.\ Phys.\  {\bf 113}, 843 (2005)
  [arXiv:hep-th/0412141].
  %%CITATION = PTPKA,113,843;%%

\bibitem{N1Star}
  C.~Vafa and E.~Witten,
  %``A Strong coupling test of S duality,''
  Nucl.\ Phys.\  B {\bf 431}, 3 (1994)
  [arXiv:hep-th/9408074];
  %%CITATION = NUPHA,B431,3;%%
  N.~Dorey,
  %``An elliptic superpotential for softly broken N = 4 supersymmetric
  %Yang-Mills theory,''
  JHEP {\bf 9907}, 021 (1999)
  [arXiv:hep-th/9906011];
  %%CITATION = JHEPA,9907,021;%%
  N.~Dorey and S.~P.~Kumar,
  %``Softly-broken N = 4 supersymmetry in the large-N limit,''
  JHEP {\bf 0002}, 006 (2000)
  [arXiv:hep-th/0001103]

 \bibitem{PS1}
  %%CITATION = JHEPA,0002,006;%%
  J.~Polchinski and M.~J.~Strassler,
  %``The string dual of a confining four-dimensional gauge theory,''
  arXiv:hep-th/0003136.
  %%CITATION = HEP-TH/0003136;%%

  \bibitem{Migdal77}
  A.~A.~Migdal,
  %``Multicolor QCD As Dual Resonance Theory,''
  Annals Phys.\  {\bf 109}, 365 (1977).
  %%CITATION = APNYA,109,365;%%

 \bibitem{Shifman05}
  M.~Shifman,
  %``Highly excited hadrons in QCD and beyond,''
  arXiv:hep-ph/0507246.
  %%CITATION = HEP-PH 0507246;%%

  \bibitem{emerging}
  J.~Erlich, G.~D.~Kribs and I.~Low,
  %``Emerging holography,''
  arXiv:hep-th/0602110;
  %%CITATION = HEP-TH 0602110;%%
    A.~Falkowski and M.~Perez-Victoria,
  %``Holography, Pade approximants and deconstruction,''
  JHEP {\bf 0702}, 086 (2007)
  [arXiv:hep-ph/0610326].
  %%CITATION = JHEPA,0702,086;%%

\bibitem{ShifmanQuarkHadronDuality}
  B.~Blok, M.~A.~Shifman and D.~X.~Zhang,
  %``An illustrative example of how quark-hadron duality might work,''
  Phys.\ Rev.\  D {\bf 57}, 2691 (1998)
  [Erratum-ibid.\  D {\bf 59}, 019901 (1999)]
  [arXiv:hep-ph/9709333];
  %%CITATION = PHRVA,D57,2691;%%
  M.~A.~Shifman,
  %``Quark-hadron duality,''
  arXiv:hep-ph/0009131.
  %%CITATION = HEP-PH/0009131;%%

 \bibitem{PQW}
  E.~C.~Poggio, H.~R.~Quinn and S.~Weinberg,
  %``Smearing The Quark Model,''
  Phys.\ Rev.\ D {\bf 13}, 1958 (1976).
  %%CITATION = PHRVA,D13,1958;%%

\bibitem{OPEtoSpectrum}
  M.~Golterman and S.~Peris,
  %``On the use of the operator product expansion to constrain the hadron
  %spectrum,''
  Phys.\ Rev.\  D {\bf 67}, 096001 (2003)
  [arXiv:hep-ph/0207060];
  %%CITATION = PHRVA,D67,096001;%%
  O.~Cata,
  %``Towards understanding Regge trajectories in holographic QCD,''
  Phys.\ Rev.\  D {\bf 75}, 106004 (2007)
  [arXiv:hep-ph/0605251];
  %%CITATION = PHRVA,D75,106004;%%
  J.~Mondejar and A.~Pineda,
  %``Constraints on Regge models from perturbation theory,''
  JHEP {\bf 0710}, 061 (2007)
  [arXiv:0704.1417 [hep-ph]];
  %%CITATION = JHEPA,0710,061;%%
  P.~Masjuan and S.~Peris,
  %``A Rational Approach to Resonance Saturation in large-Nc QCD,''
  JHEP {\bf 0705}, 040 (2007)
  [arXiv:0704.1247 [hep-ph]].
  %%CITATION = JHEPA,0705,040;%%

 \bibitem{CoulombGauge}
   A.~P.~Szczepaniak and E.~S.~Swanson,
  %``Coulomb gauge QCD, confinement, and the constituent representation,''
  Phys.\ Rev.\  D {\bf 65}, 025012 (2002)
  [arXiv:hep-ph/0107078];
  %%CITATION = PHRVA,D65,025012;%%
    A.~P.~Szczepaniak and E.~S.~Swanson,
  %``The low lying glueball spectrum,''
  Phys.\ Lett.\  B {\bf 577}, 61 (2003)
  [arXiv:hep-ph/0308268];
  %%CITATION = PHLTA,B577,61;%%
  N.~Ligterink and E.~S.~Swanson,
  %``A Coulomb gauge model of mesons,''
  Phys.\ Rev.\  C {\bf 69}, 025204 (2004)
  [arXiv:hep-ph/0310070].
  %%CITATION = PHRVA,C69,025204;%%

  \bibitem{StaticPotentialAdSCFT}
  S.~J.~Rey and J.~T.~Yee,
  %``Macroscopic strings as heavy quarks in large N gauge theory and  anti-de
  %Sitter supergravity,''
  Eur.\ Phys.\ J.\  C {\bf 22}, 379 (2001)
  [arXiv:hep-th/9803001];
  %%CITATION = EPHJA,C22,379;%%
    J.~M.~Maldacena,
  %``Wilson loops in large N field theories,''
  Phys.\ Rev.\ Lett.\  {\bf 80}, 4859 (1998)
  [arXiv:hep-th/9803002].
  %%CITATION = PRLTA,80,4859;%%

 \bibitem{Zwanziger}
   D.~Zwanziger,
  %``No confinement without Coulomb confinement,''
  Phys.\ Rev.\ Lett.\  {\bf 90}, 102001 (2003)
  [arXiv:hep-lat/0209105].
  %%CITATION = PRLTA,90,102001;%%

\bibitem{BFBF}
  H.~Boschi-Filho, N.~R.~F.~Braga and C.~N.~Ferreira,
  %``Static strings in Randall-Sundrum scenarios and the quark anti-quark
  %potential,''
  Phys.\ Rev.\  D {\bf 73}, 106006 (2006)
  [Erratum-ibid.\  D {\bf 74}, 089903 (2006)]
  [arXiv:hep-th/0512295];
  %%CITATION = PHRVA,D73,106006;%%
  H.~Boschi-Filho, N.~R.~F.~Braga and C.~N.~Ferreira,
  %``Static strings in Randall-Sundrum scenarios and the quark anti-quark
  %potential: Erratum,''
  arXiv:hep-th/0610131.
  %%CITATION = HEP-TH/0610131;%%

\bibitem{SoftMetric}
  O.~Andreev,
  %``1/q**2 corrections and gauge / string duality,''
  Phys.\ Rev.\  D {\bf 73}, 107901 (2006)
  [arXiv:hep-th/0603170].
  %%CITATION = PHRVA,D73,107901;%%

\bibitem{AdSQCDPotential}
    O.~Andreev and V.~I.~Zakharov,
  %``Heavy-quark potentials and AdS/QCD,''
  Phys.\ Rev.\  D {\bf 74}, 025023 (2006)
  [arXiv:hep-ph/0604204].
  %%CITATION = PHRVA,D74,025023;%%
    C.~D.~White,
  %``The Cornell potential from general geometries in AdS / QCD,''
  Phys.\ Lett.\  B {\bf 652}, 79 (2007)
  [arXiv:hep-ph/0701157].
  %%CITATION = PHLTA,B652,79;%%
    F.~Giannuzzi,
  %``\eta_b and \eta_c radiative decays in the Salpeter model with the AdS/QCD
  %inspired potential,''
  arXiv:0810.2736 [hep-ph].
  %%CITATION = ARXIV:0810.2736;%%

  \bibitem{sphericity}
  J.~D.~Bjorken and S.~J.~Brodsky,
  %``Statistical Model For Electron-Positron Annihilation Into Hadrons,''
  Phys.\ Rev.\  D {\bf 1}, 1416 (1970).
  %%CITATION = PHRVA,D1,1416;%%

  \bibitem{softwall}
  A.~Karch, E.~Katz, D.~T.~Son and M.~A.~Stephanov,
  %``Linear confinement and AdS/QCD,''
  Phys.\ Rev.\  D {\bf 74}, 015005 (2006)
  [arXiv:hep-ph/0602229].
  %%CITATION = PHRVA,D74,015005;%%

\bibitem{moresoftwall}
    C.~Cs\'aki and M.~Reece,
  %``Toward a systematic holographic QCD: A braneless approach,''
  JHEP {\bf 0705}, 062 (2007)
  [arXiv:hep-ph/0608266].
  %%CITATION = JHEPA,0705,062;%%
    U.~Gursoy and E.~Kiritsis,
  %``Exploring improved holographic theories for QCD: Part I,''
  JHEP {\bf 0802}, 032 (2008)
  [arXiv:0707.1324 [hep-th]].
  %%CITATION = JHEPA,0802,032;%%
  U.~Gursoy, E.~Kiritsis and F.~Nitti,
  %``Exploring improved holographic theories for QCD: Part II,''
  JHEP {\bf 0802}, 019 (2008)
  [arXiv:0707.1349 [hep-th]].
  %%CITATION = JHEPA,0802,019;%%
  B.~Batell and T.~Gherghetta,
  %``Dynamical Soft-Wall AdS/QCD,''
  Phys.\ Rev.\  D {\bf 78}, 026002 (2008)
  [arXiv:0801.4383 [hep-ph]].
  %%CITATION = PHRVA,D78,026002;%%
  W.~de Paula, T.~Frederico, H.~Forkel and M.~Beyer,
  %``Dynamical AdS/QCD with area-law confinement and linear Regge
  %trajectories,''
  arXiv:0806.3830 [hep-ph].
  %%CITATION = ARXIV:0806.3830;%%

  \bibitem{KatzOkui}
    E.~Katz and T.~Okui,
  %``The 't Hooft Model As A Hologram,''
  arXiv:0710.3402 [hep-th].
  %%CITATION = ARXIV:0710.3402;%%

 \bibitem{BrodskyDeTeramond}
   S.~J.~Brodsky and G.~F.~de Teramond,
  %``Light-front hadron dynamics and AdS/CFT correspondence,''
  Phys.\ Lett.\  B {\bf 582}, 211 (2004)
  [arXiv:hep-th/0310227];
  %%CITATION = PHLTA,B582,211;%%
  G.~F.~de Teramond and S.~J.~Brodsky,
  %``The hadronic spectrum of a holographic dual of QCD,''
  Phys.\ Rev.\ Lett.\  {\bf 94}, 201601 (2005)
  [arXiv:hep-th/0501022];
  %%CITATION = PRLTA,94,201601;%%
  S.~J.~Brodsky and G.~F.~de Teramond,
  %``Hadronic spectra and light-front wavefunctions in holographic QCD,''
  Phys.\ Rev.\ Lett.\  {\bf 96}, 201601 (2006)
  [arXiv:hep-ph/0602252].
  %%CITATION = PRLTA,96,201601;%%

  \bibitem{OldRegge}
    G.~F.~Chew and S.~C.~Frautschi,
  %``REGGE TRAJECTORIES AND THE PRINCIPLE OF MAXIMUM STRENGTH FOR STRONG
  %INTERACTIONS,''
  Phys.\ Rev.\ Lett.\  {\bf 8}, 41 (1962);
  %%CITATION = PRLTA,8,41;%%
    J.~R.~Forshaw and D.~A.~Ross,
  %``Quantum chromodynamics and the pomeron,''
  Cambridge Lect.\ Notes Phys.\  {\bf 9}, 1 (1997).
  %%CITATION = 00385,9,1;%%

  \bibitem{LatticeStrings}
    K.~J.~Juge, J.~Kuti and C.~Morningstar,
  %``Fine structure of the QCD string spectrum,''
  Phys.\ Rev.\ Lett.\  {\bf 90}, 161601 (2003)
  [arXiv:hep-lat/0207004].
  %%CITATION = PRLTA,90,161601;%%

  \bibitem{Casher:1978wy}
  A.~Casher, H.~Neuberger and S.~Nussinov,
  %``Chromoelectric Flux Tube Model Of Particle Production,''
  Phys.\ Rev.\  D {\bf 20}, 179 (1979).
  %%CITATION = PHRVA,D20,179;%%

 \bibitem{Glendenning:1983qq}
   N.~K.~Glendenning and T.~Matsui,
  %``Creation Of Anti-Q Q Pair In A Chromoelectric Flux Tube,''
  Phys.\ Rev.\  D {\bf 28}, 2890 (1983).
  %%CITATION = PHRVA,D28,2890;%%

  \bibitem{Gurvich:1979nq}
  E.~G.~Gurvich,
  %``The Quark Anti-Quark Pair Production Mechanism In A Quark Jet,''
  Phys.\ Lett.\  B {\bf 87}, 386 (1979).
  %%CITATION = PHLTA,B87,386;%%

  \bibitem{Isgur:1983wj}
  N.~Isgur and J.~E.~Paton,
  %``A Flux Tube Model For Hadrons,''
  Phys.\ Lett.\  B {\bf 124}, 247 (1983);
  %%CITATION = PHLTA,B124,247;%%
  N.~Isgur and J.~E.~Paton,
  %``A Flux Tube Model For Hadrons In QCD,''
  Phys.\ Rev.\  D {\bf 31}, 2910 (1985);
  %%CITATION = PHRVA,D31,2910;%%
  R.~Kokoski and N.~Isgur,
  %``Meson Decays By Flux Tube Breaking,''
  Phys.\ Rev.\  D {\bf 35}, 907 (1987).
  %%CITATION = PHRVA,D35,907;%%

  \bibitem{Andersson:1983ia}
  B.~Andersson, G.~Gustafson, G.~Ingelman and T.~Sjostrand,
  %``Parton Fragmentation And String Dynamics,''
  Phys.\ Rept.\  {\bf 97}, 31 (1983).
  %%CITATION = PRPLC,97,31;%%

  \bibitem{GluonChain}
  J.~Greensite and C.~B.~Thorn,
  %``Gluon chain model of the confining force,''
  JHEP {\bf 0202}, 014 (2002)
  [arXiv:hep-ph/0112326].
  %%CITATION = JHEPA,0202,014;%%

\bibitem{Greensite:2003bk}
  J.~Greensite,
  %``The confinement problem in lattice gauge theory,''
  Prog.\ Part.\ Nucl.\ Phys.\  {\bf 51}, 1 (2003)
  [arXiv:hep-lat/0301023].
  %%CITATION = PPNPD,51,1;%%

\bibitem{Zwiebach:2004tj}
  B.~Zwiebach,
  {\em A First Course in String Theory.}
%\href{http://www.slac.stanford.edu/spires/find/hep/www?irn=6000347}{SPIRES entry}
Cambridge, UK: Univ. Pr. (2004) 558 p

\bibitem{Quirks}
  J.~Kang and M.~A.~Luty,
  %``Macroscopic Strings and 'Quirks' at Colliders,''
  arXiv:0805.4642 [hep-ph].
  %%CITATION = ARXIV:0805.4642;%%

\bibitem{bubbles}
  S.~R.~Coleman,
  %``The Fate Of The False Vacuum. 1. Semiclassical Theory,''
  Phys.\ Rev.\  D {\bf 15} (1977) 2929
  [Erratum-ibid.\  D {\bf 16} (1977) 1248];
  %%CITATION = PHRVA,D15,2929;%%
    C.~G.~.~Callan and S.~R.~Coleman,
  %``The Fate Of The False Vacuum. 2. First Quantum Corrections,''
  Phys.\ Rev.\  D {\bf 16} (1977) 1762;
  %%CITATION = PHRVA,D16,1762;%%
  S.~R.~Coleman, ``The Uses of Instantons", in
{\em Aspects of Symmetry: Selected Erice Lectures} (Cambridge University Press, Cambridge, 1985).

\bibitem{fluxbubble}
  A.~Casher, H.~Neuberger and S.~Nussinov,
  %``Multiparticle Production By Bubbling Flux Tubes,''
  Phys.\ Rev.\  D {\bf 21} (1980) 1966.
  %%CITATION = PHRVA,D21,1966;%%

  \bibitem{barger}
  V.~Barger and R.~Phillips, p. 184 {\em Collider Physics} (Westview Press, Boulder CO, 1991).

\end{thebibliography}
\end{document}